\documentclass[a4paper,fleqn,usenatbib]{mnras}

\usepackage{newtxtext,newtxmath}
\usepackage[T1]{fontenc}
\usepackage{ae,aecompl}

\usepackage{graphicx}	% Including figure files
\usepackage{amsmath}	% Advanced maths commands
\usepackage{amssymb}	% Extra maths symbols

\usepackage{ulem}

\newcommand{\epichj}{K2-260}
\newcommand{\epicws}{K2-261}
\newcommand{\feh}{\ensuremath{\left[{\rm Fe}/{\rm H}\right]}}  % [Fe/H]
\newcommand{\teff}{\ensuremath{T_{\rm eff}}}
\newcommand{\loggstar}{\ensuremath{\log{g_{\star}}}}
\newcommand{\vsinistar}{\ensuremath{v\sin{i_{\star}}}}
\newcommand{\msun}{\ensuremath{\,M_{\odot}}}
\newcommand{\rsun}{\ensuremath{\,R_{\odot}}}

\newcommand{\mj}{\ensuremath{\,M_{\rm J}}}
\newcommand{\rj}{\ensuremath{\,R_{\rm J}}}

\newcommand{\re}{\ensuremath{\,R_{\oplus}}}

\newcommand{\kms}{\ensuremath{\rm km\ s^{-1}}}
\newcommand{\ms}{\ensuremath{\rm m\ s^{-1}}}
\newcommand{\mstar}{\ensuremath{M_{\star}}}
\newcommand{\rstar}{\ensuremath{R_{\star}}}
\newcommand{\kepler}{{\it Kepler}}

\title[A hot Jupiter and a warm Saturn from K2]{K2-260~b: a hot Jupiter transiting an F star, and K2-261~b: a warm Saturn around a bright G star}

\author[M. C. Johnson et al.]{
M. C. Johnson,$^{1}$\thanks{E-mail: johnson.7240@osu.edu (MCJ)}
F. Dai,$^{2,3}$ %light curves, detrending, secondary eclipse fitting &c
A. B. Justesen,$^{4}$ %spectral type
D. Gandolfi,$^{5}$ %RVs
A. P. Hatzes,$^{6}$ %frequency analysis
G. Nowak,$^{7,8}$ %HARPS-N RVs for C13_1830, planned R-M, much discussion
\newauthor
M. Endl,$^{9}$ %McDonald data/Kea parameters
W. D. Cochran,$^{9}$ %McDonald data/Kea parameters, UVW discussion
D. Hidalgo,$^{7,8}$ %EVEREST light curve, search for additional transits
N. Watanabe,$^{10}$ %MuSCAT2 observer, reduction
H. Parviainen,$^{7,8}$ %MuSCAT2 light curves
\newauthor
T. Hirano,$^{11,12}$ %IRCS observations and analysis
S. Villanueva Jr.,$^{1,13}$ %DEMONEXT light curves
J. Prieto-Arranz,$^{7,8}$ %FastCam observations
N. Narita,$^{14,15,16,8,17}$ %Arranged MuSCAT2 observations
E. Palle,$^{7,8}$ %arranged MuSCAT2 observations
\newauthor
E. W. Guenther,$^{6}$ %FIES observations over xmas
O. Barrag\'an,$^{5}$ %HARPS observations
T. Trifonov,$^{18}$ %HARPS observations via time exchange
P. Niraula,$^{19}$ %phase curve analysis
P. J. MacQueen,$^{9}$ %Recon spectrum of C14_8078
\newauthor
J. Cabrera,$^{20}$
Sz. Csizmadia,$^{20}$
Ph. Eigm\"uller,$^{20}$
S. Grziwa,$^{21}$
J. Korth,$^{21}$
M. P\"atzold,$^{21}$
\newauthor
A. M. S. Smith,$^{20}$
S. Albrecht,$^{4}$
R. Alonso,$^{7,8}$
H. Deeg,$^{7,8}$
A. Erikson,$^{20}$
M. Esposito,$^{6}$
\newauthor
M. Fridlund,$^{22,23}$
A. Fukui,$^{15}$
N. Kusakabe,$^{15,16}$
M. Kuzuhara,$^{15,16}$
J. Livingston,$^{24}$
\newauthor
P. Monta\~nes Rodriguez,$^{7,8}$
D. Nespral,$^{7,8}$
C. M. Persson,$^{22}$
T. Purismo,$^{25}$
S. Raimundo,$^{26}$
\newauthor
H. Rauer,$^{20,27}$
I. Ribas,$^{28}$
M. Tamura,$^{14,15,16}$
V. Van Eylen,$^{23}$
and J. N. Winn$^{3}$
\\
$^{1}$Department of Astronomy, The Ohio State University, 140 West 18th Ave., Columbus, OH 43210, USA\\
$^{2}$Department of Physics and Kavli Institute for Astrophysics and Space Research, Massachusetts Institute of Technology, Cambridge, MA, \\
02139, USA \\
$^{3}$Department of Astrophysical Sciences, Princeton University, 4 Ivy Lane, Princeton, NJ, 08544, USA \\
$^{4}$Stellar Astrophysics Centre, Department of Physics and Astronomy, Aarhus University, Ny Munkegrade 120, DK-8000 Aarhus C, Denmark \\
$^{5}$Dipartimento di Fisica, Universit\`a di Torino, Via P. Giuria 1, I-10125 Torino, Italy \\
$^{6}$Th\"uringer Landessternwarte Tautenburg, Sternwarte 5, D-07778 Tautenberg, Germany \\
$^{7}$Departamento de Astrof\'isica, Universidad de La Laguna, E-38206, Tenerife, Spain \\
$^{8}$Instituto de Astrof\'isica de Canarias, C/ V\'ia L\'actea s/n, E-38205, La Laguna, Tenerife, Spain \\
$^{9}$McDonald Observatory, University of Texas at Austin, 2515 Speedway, Stop C1400, Austin, TX 78712, USA \\
$^{10}$Department of Astronomical Science, Graduate University for Advanced Studies (SOKENDAI), Mitaka, Tokyo 181-8588, Japan \\
$^{11}$Department of Earth and Planetary Sciences, Tokyo Institute of Technology, 2-12-1 Ookayama, Meguro-ku, Tokyo 152-8551, Japan \\
$^{12}$Institute for Astronomy, University of Hawaii, 2680 Woodlawn Drive, Honolulu, Hawaii 96822, USA \\
$^{13}$NSF Graduate Research Fellow \\
$^{14}$Department of Astronomy, The University of Tokyo, 7-3-1 Hongo, Bunkyo-ku, Tokyo 113-0033, Japan \\
$^{15}$National Astronomical Observatory of Japan, NINS, 2-21-1 Osawa, Mitaka, Tokyo 181-8588, Japan \\
$^{16}$Astrobiology Center, NINS, 2-21-1 Osawa, Mitaka, Tokyo 181-8588, Japan\\
$^{17}$JST, PRESTO, 7-3-1 Hongo, Bunkyo-ku, Tokyo 113-0033, Japan\\
$^{18}$Max-Planck-Institut f\"ur Astronomie, K\"onigstuhl 17, D-69117 Heidelberg, Germany\\
$^{19}$Astronomy Department and Van Vleck Observatory, Wesleyan University, Middletown, CT 06459, USA \\
$^{20}$Institute of Planetary Research, German Aerospace Center, Rutherfordstrasse 2, 12489 Berlin, Germany \\
$^{21}$Rheinisches Institut f\"ur Umweltforschung, Abteilung Planetenforschung an der Universit\"at zu K\"oln, Aachener Strasse 209, 50931 K\"oln, \\
Germany \\
$^{22}$Department of Space, Earth and Environment, Chalmers University of Technology, Onsala Space Observatory, 439 92 Onsala, Sweden \\
$^{23}$Leiden Observatory, University of Leiden, PO Box 9513, 2300 RA, Leiden, The Netherlands \\
$^{24}$Department of Astronomy, Graduate School of Science, The University of Tokyo, Hongo 7-3-1, Bunkyo-ku, Tokyo, 113-0033, Japan \\
$^{25}$Nordic Optical Telescope, Rambla Jos\'e Ana Fern\'andez P\'erez 7, 38711 Bre\~na Baja, Spain \\
$^{26}$Dark Cosmology Centre, Niels Bohr Institute, University of Copenhagen, Denmark \\
$^{27}$Center for Astronomy and Astrophysics, TU Berlin, Hardenbergstr. 36, 10623 Berlin, Germany \\
$^{28}$Institut de Ci\`encies de l'Espai (CSIC-IEEC), Carrer de Can Magrans, Campus UAB, E-08193 Bellaterra, Spain \\
\\
\\
\\
\\
\\
\\
\\
\\
\\
\\
}

\date{Accepted 2018 August 9. Received 2018 August 8; in original form 2018 April 4}

\pubyear{2018}

\begin{document}
\label{firstpage}
\pagerange{\pageref{firstpage}--\pageref{lastpage}}
\maketitle

\newpage

\begin{abstract}
We present the discovery and confirmation of two new transiting giant planets from the \kepler\ extended mission {\it K2}. \epichj~b is a hot Jupiter transiting a $V=12.7$ F6V star in {\it K2} Field 13, with a mass and radius of \mstar$=1.39_{-0.06}^{+0.05}$ \msun\ and \rstar$=1.69 \pm 0.03$ \rsun. The planet has an orbital period of $P=2.627$ days, and a mass and radius of $M_P=1.42^{+0.31}_{-0.32}$ \mj\ and $R_P=1.552^{+0.048}_{-0.057}$ \rj. 
This is the first {\it K2} hot Jupiter with a detected secondary eclipse in the \kepler\ bandpass, with a depth of $71 \pm 15$ ppm, which we use to estimate a geometric albedo of $A_g\sim0.2$. We also detected a candidate stellar companion at 0.6'' from \epichj; we find that it is very likely physically associated with the system, in which case it would be an M5-6V star at a projected separation of $\sim400$ AU. \epicws~b is a warm Saturn transiting a bright ($V=10.5$) G7IV/V star in {\it K2} Field 14. The host star is a metal-rich (\feh$=0.36 \pm 0.06$), mildly evolved $1.10_{-0.02}^{+0.01}$ \msun\ star with \rstar$=1.65 \pm 0.04$ \rsun. Thanks to its location near the main sequence turn-off, we can measure a relatively precise age of $8.8_{-0.3}^{+0.4}$ Gyr. The planet has $P=11.633$ days, $M_P=0.223 \pm 0.031$ \mj, and $R_P=0.850^{+0.026}_{-0.022}$ \rj, and its orbit is eccentric ($e=0.39 \pm 0.15$). Its brightness and relatively large transit depth make this one of the best known warm Saturns for follow-up observations to further characterize the planetary system.

\end{abstract}


%%%%%%%%%%%%%%%%%%%%%%%%%%%%%%%%%%%%%%%%%%%%%%%%%%

%%%%%%%%%%%%%%%%% BODY OF PAPER %%%%%%%%%%%%%%%%%%

\section{Introduction}

Our knowledge of exoplanets has been revolutionised by the \kepler\ mission \citep{Borucki10}, which during its prime mission discovered thousands of exoplanet candidates and confirmed exoplanets, providing powerful statistical measurements of planetary populations. After the failure of a second reaction wheel  in 2013, the \kepler\ spacecraft began the {\it K2} extended mission \citep{Howell14}, pointing at a succession of fields around the ecliptic for $\sim80$ days per field. The {\it K2} mission has resulted in the detection of over 200 confirmed and validated planets, and even more planet candidates, to date \citep[e.g.,][]{Mayo18}. On average {\it K2} target stars are brighter than those from the \kepler\ prime mission, resulting in a population of planets that is more amenable to direct confirmation, and further observations to characterize these systems in more detail. 

Building a large sample of planets that are characterized in detail will help to move the field of exoplanet population statistics beyond the parameters that have been probed by \kepler\ \citep[principally planetary radius and orbital period; e.g.,][]{Burke15}. Large statistical studies of planetary atmospheres, spin-orbit misalignments, and other properties measurable via follow-up observations promise to set much more powerful constraints upon models of planet formation, migration, and evolution, and illuminate planetary astrophysics in general \citep[e.g.,][]{MortonJohnson11,Sing16}.

Although a large number of planets and planet candidates around bright stars will soon be provided by the {\it Transiting Exoplanet Survey Satellite} \citep[{\it TESS};][]{Ricker15} mission, {\it K2} planets are complementary to those that will be found by {\it TESS}. {\it K2} observes in the ecliptic plane, which will not be covered by {\it TESS} during its prime mission. {\it K2} is thus helping {\it TESS} to complete an all-sky catalogue of transiting short-period planets around relatively bright stars. Additionally, as {\it K2} campaigns are significantly longer than {\it TESS} pointings ($\sim80$ versus $\sim27$ days), {\it K2} is capable of finding longer-period planets than {\it TESS} typically will over most of the sky. 

We present here the discovery, confirmation, and additional characterization of two new transiting planets from {\it K2}: \epichj~b and \epicws~b. \epichj~b is a hot Jupiter transiting a $V=12.7$ mid-F star, while \epicws~b is a warm Saturn transiting a $V=10.6$ late-G star near the main sequence turn-off. Both are amenable to further observations, and can add to the number of giant planets characterized in detail.

These planets were discovered and characterized as part of the KESPRINT collaboration to find planets using {\it K2}. Our team has confirmed and characterized over two dozen transiting planets using {\it K2}, including hot Jupiters \citep[e.g.,][]{Grziwa16,Hirano16,K2HotJupiters}, longer-period giant planets \citep[e.g.,][]{Smith17,VanEylen18}, ultra-short-period planets \citep[e.g.,][]{Dai17,Smith18}, multiplanet systems \citep[e.g.,][]{Gandolfi17,Niraula17}, as well as other planets \citep[e.g.,][]{Narita17}.

\section{Observations}

\subsection{{\it K2} Photometry}

The \kepler\ spacecraft observed {\it K2} Field 13 from 2017 March 8 to May 27 UT, a span of 79.0 days. Field 13 is centred at RA$=04^h51^m11^s$, Dec$=+20^{\circ}47'11''$ (J2000.0). The star EPIC 246911830 (\epichj) was proposed for observations by programs GO13122 (P.I.\ Howard) and GO13071 (P.I.\ Charbonneau). {\it K2} Field 14 was observed for 70.4 days between 2017 May 31 and August 9 UT. Field 14 is centred at RA$=10^h42^m44^s$, Dec$=+06^{\circ}51'06''$ (J2000.0). EPIC 201498078 (\epicws) was proposed for observations by programs GO14009 (P.I.\ Charbonneau), GO14020 (P.I.\ Adams), GO14021 (P.I.\ Howard), and GO14028 (P.I.\ Cochran).

 We conducted three parallel searches of all of the light curves from Campaigns 13 and 14 for transits using three separate methodologies. 
One method utilizes aperture photometry to extract light curves from the {\it K2} raw pixel data, and decorrelates the light curves based upon the centroid motion to account for the roll of the spacecraft, as described in \cite{Dai16K2}. We then perform a standard BLS \citep[Box Least Squares;][]{Kovacs02} transit search upon the light curves.

The other two methods both used the {\it K2} light curves for Campaigns 13 and 14 as extracted and corrected for systematics per \cite{VanderburgJohnson15} as provided by A.\ Vanderburg\footnote{\url{https://www.cfa.harvard.edu/~avanderb/k2.html}}. We used the \textsc{exotrans} package \citep{Grziwa12} in order to search the light curves for transits. This package utilizes the \textsc{varlet} \citep{GrziwaPaetzold16} wavelet-based filtering technique in order to remove stellar and systematic variability from the light curves, along with a BLS transit search algorithm. If a transit is detected, it is removed using another wavelet-based filtering technique, \textsc{phalet}, and the light curve searched again for additional transits using BLS.

We also used the code \textsc{dst} \citep{Cabrera12} to search for transits in the \cite{VanderburgJohnson15} light curves. This algorithm is conceptually similar to BLS, but uses a more realistic transit shape rather than an inverted tophat for the transit search, accounts for the potential presence of transit timing variations, and uses improved statistical methods. 

\begin{figure*}
	\includegraphics[width=\textwidth]{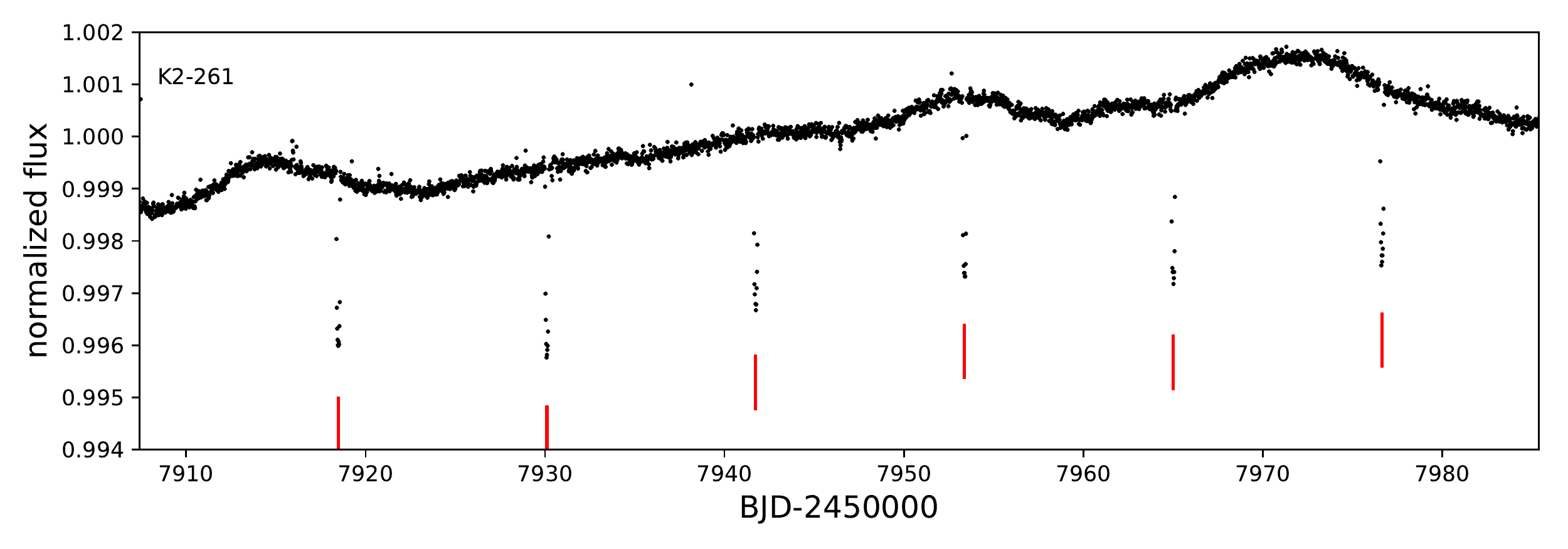}\\
    \includegraphics[width=\textwidth]{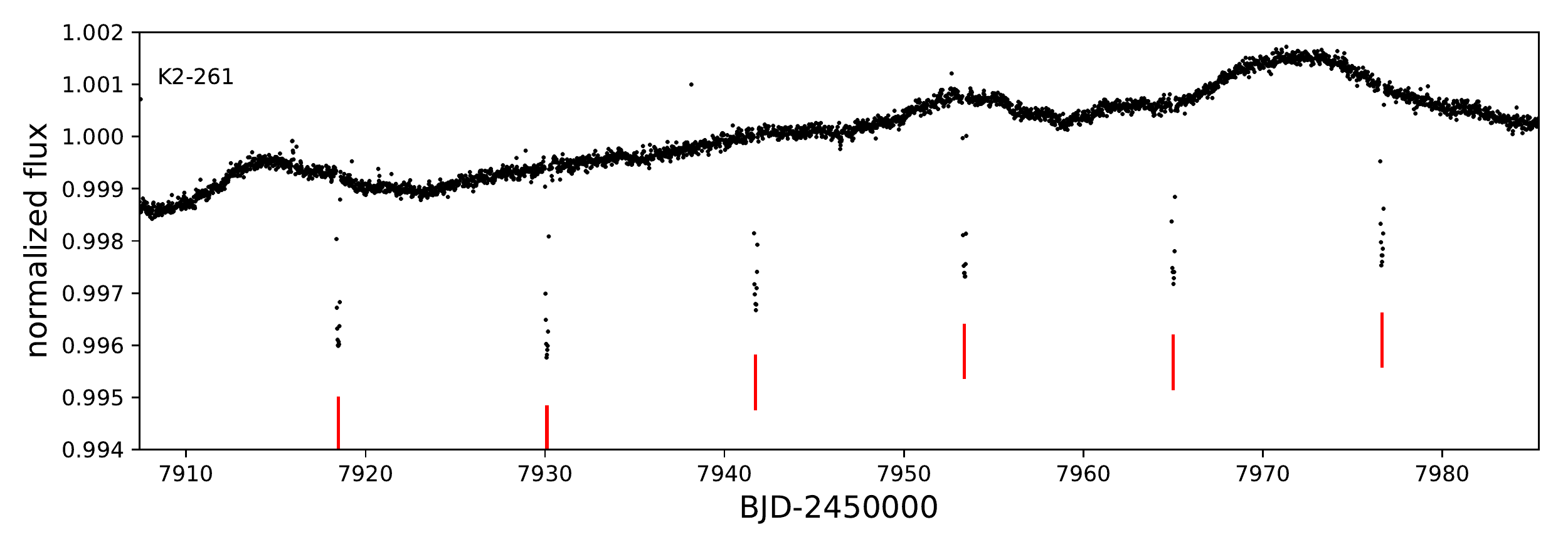}\\
    \caption{{\it K2} light curves of \epichj\ (top), and \epicws\ (bottom), as detrended and corrected for systematics by Vanderburg \& Johnson (2015). Note the rotational variability of \epichj, which is discussed in \S\ref{sec:stellarrotation}. For both systems we remove the stellar and instrumental variations prior to fitting the transits. 
    Vertical bars mark the locations of each transit.} 
    \label{fig:K2LCs}
\end{figure*}

All three search techniques detected significant transits for both \epichj\ (with a period of $\sim2.6$ days and a depth of $\sim1$ per cent) and \epicws\ (with a period of $\sim11.6$ days and a depth of $\sim0.3$ per cent); 
indeed, the transits are easily visible by eye in the detrended {\it K2} light curves (Fig.~\ref{fig:K2LCs}). None of the algorithms identified any additional candidate transits for either star.

We list the coordinates, magnitudes, and other identifying information for \epichj\ and \epicws\ in Table~\ref{tab:litpars}.

\begin{table*}
	\centering
	\caption{Coordinates, magnitudes, and kinematics of \epichj\ and \epicws}
	\label{tab:litpars}
	\begin{tabular}{lccr} 
		\hline
		Parameter & \epichj & \epicws & Reference \\
		\hline
		\multicolumn{4}{p{300pt}}{Coordinates and Identifiers} \\
        RA (J2000.0) &  	$05^h07^m28^s.16$ & $10^h52^m07^s.78$ & 1 \\
        Dec (J2000.0) & $+16^{\circ}52'03''.78$ &  	$+00^{\circ}29'36''.07$ & 1 \\
        EPIC & 246911830 & 201498078 \\ 
        TYC & &  255-257-1 & \\
        UCAC4	&  535-011514 &  453-051151 & \\
        2MASS	&  J05072816+1652037 &  J10520778+0029359 	& \\
        AllWISE & J050728.15+165203.7 & J105207.76+002935.6 & \\
        \hline
        \multicolumn{4}{p{300pt}}{Magnitudes} \\
        $Kp$ (mag) & $12.465$ & $10.451$ & 1 \\
        $G$ (mag) & $12.467$ & $10.459$ & 2 \\
        $Bp$ (mag) & $12.798$ & $10.872$ & 2 \\
        $Rp$ (mag) & $11.973$ & $9.917$ & 2 \\
        $B$ (mag) & $13.217 \pm	0.042$ & $11.561 \pm	0.086$ & 1 \\
        $V$ (mag) & $12.69 \pm	0.11$ & $10.612 \pm	0.059$ & 1 \\
        $g$ (mag) & $12.881 \pm	0.010$ & $10.979 \pm	0.010 	$ & 1 \\
        $r$ (mag) & $12.449 \pm	0.030$ & $10.402 \pm	0.020$ & 1 \\
        $i$ (mag) & $12.287 \pm	0.010$ & $10.226 \pm	0.020$ & 1 \\
        $J$ (mag) & $11.400 \pm	0.023$ & $9.337 \pm	0.030$ & 1 \\
        $H$ (mag) & $11.189 \pm	0.032$ & $8.920 \pm	0.042$ & 1 \\
        $K$ (mag) & $11.093 \pm	0.021$ & $8.890 \pm	0.022$ & 1 \\
        $W1$ (mag) & $11.039 \pm	0.023$ & $8.828 \pm	0.023$ & 3 \\
        $W2$ (mag) & $11.036 \pm	0.021$ & $8.897 \pm	0.020$ & 3 \\
        $W3$ (mag) & $10.895 \pm	0.129$ & $8.819 \pm	0.031$ & 3 \\
        $W4$ (mag) & $>9.006$ 	& $>8.281$ 	 & 3 \\
        $B-V$ (mag) & $0.53 \pm 0.12$ & $0.95 \pm 0.10$ & 4 \\
        $J-K$ (mag) & $0.307 \pm 0.031$ & $0.447 \pm 0.037$ & 4 \\
		\hline
        \multicolumn{4}{p{300pt}}{Distance and Velocities} \\
        $\mu_{\mathrm{RA}}$ (mas yr$^{-1}$) & $0.667 \pm 0.078$ & $-23.664 \pm	0.075$ & 2 \\
        $\mu_{\mathrm{Dec}}$ (mas yr$^{-1}$) & $-6.045 \pm 0.051$ & $-44.171 \pm 0.068$ & 2 \\
        $\Pi$ (mas) & $1.479 \pm 0.042$ & $4.660 \pm 0.043$ & 2 \\
        $d$ (pc) & $676 \pm 19$ & $214.6 \pm 2.0$ & 2 \\
        $v_{\mathrm{rad}}$ (\kms) & $29.1 \pm 2.7$ & $3.28 \pm 0.52$ & 2 \\
        $U$ (\kms) & $15.8 \pm 2.6$ & $-8.99 \pm 0.13$ & This Work \\
        $V$ (\kms) & $-6.15 \pm 0.57$ & $-29.14 \pm 0.49$ & This Work \\
        $W$ (\kms) & $-9.46 \pm 0.73$ & $-21.81 \pm 0.50$ & This Work \\
        \hline
        \multicolumn{4}{p{300pt}}{Notes: the mean stellar radial velocity $v_{\mathrm{rad}}$ is measured from our FIES spectra for \epichj\ and from our McDonald reconnaissance spectrum for \epicws. The quoted $UVW$ velocities are in the LSR frame and use the LSR definition and $UVW$ frame of \cite{Coskunoglu11}. References: 1: Ecliptic Plane Input Catalog (EPIC: \citealt{Huber16}; \url{https://archive.stsci.edu/k2/epic/search.php}); 2: {\it Gaia} DR2 \citep{Gaia18}; 3: AllWISE \citep{Cutri13}; 4: calculated from literature values given earlier in this table.}
	\end{tabular}
\end{table*}

\subsection{High-resolution imaging}
\label{sec:AO}

In order to test for the presence of nearby stars which could dilute the transit depth and perhaps indicate that the planet candidate could be a false positive, we obtained high-resolution imaging observations of both \epichj\ and \epicws.

We obtained high-resolution images of \epichj\ using the Infrared Camera and Spectrograph
\citep[IRCS:][]{Kobayashi00} with the adaptive optics (AO) system AO188: \citep[][]{Hayano10}
mounted on the 8.2~m Subaru telescope on 2018 March 29 UT. For the AO imaging, we adopted the fine-sampling 
mode (20 mas per pixel) and $H-$band filter. 
Short exposure, unsaturated frames (0.6 s $\times$ 3 coadds for each position) were first secured with five-point dithering, 
which are used for the absolute flux calibration. We then ran long-exposure sequences (15 s $\times$ 3 coadds
for each position) adopting the same five-point dithering, resulting in saturated frames of \epichj\
to look for faint nearby companions. The total exposure time amounted to 450 s for the saturated frames. 

Following \citet{Hirano16b}, we applied dark subtraction, bad pixel interpolation, flat fielding, 
and distortion correction to the raw IRCS frames. The unsaturated and saturated frames were then respectively
aligned and combined to obtain the final reduced images. 
The combined unsaturated image of \epichj\ showed that the AO-corrected full width at half maximum 
(FWHM) of the target's point spread function was $0\farcs10$. 

\begin{figure}
	\includegraphics[width=\columnwidth, trim = 0in 0in 5.5in 9in]{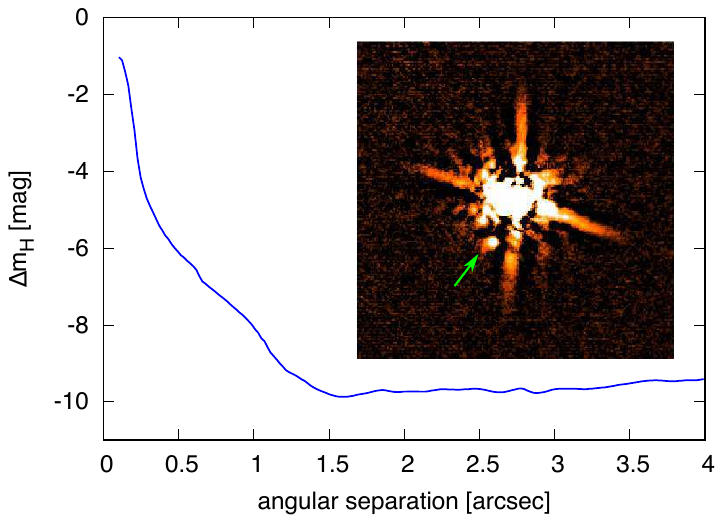}\\
    \caption{$H$-band $5\sigma$ contrast curve of \epichj\ from IRCS. Inset: high-pass filtered image. The candidate companion is marked with an arrow.}
    \label{fig:Subaru}
\end{figure}

Visual inspection of the combined saturated image suggested a presence of a faint source 
in the south-east of \epichj\, with a separation of $\sim 0\farcs6$ (see the inset of Figure~\ref{fig:Subaru}). 
In order to estimate the contrast of this candidate companion (CC), we performed aperture photometry for \epichj\ and its CC with its aperture radius being the FWHM of the target. 
For the CC's photometry, we applied high-pass filtering (with a width of $4\times\mathrm{FWHM}$) to the saturated image 
to suppress the halo of the primary star.

High-pass filtering suppresses the halo of the primary star, but it also reduces the CC's flux. In order to
estimate the flux loss from the high-pass filtering, we implemented a simulation in which an artificial companion's
signal is injected into the saturated image with the same angular separation (but with different position angles). 
The mock data were analysed using the same steps above to measure the artificial companion's flux. 
As a result, we found that approximately 25\% of the injected signal was lost by high-pass filtering. 
Taking into account this flux loss, we measured the CC's magnitude and its position. Table \ref{tab:companion}
summarizes the measured CC properties. 
To evaluate the detection level of the CC, we also drew the 5$\sigma$ contrast curve based on the scatter 
of the flux count in the annulus as a function of angular separation from \epichj. Figure~\ref{fig:Subaru} implies 
that at the separation of $\sim 0\farcs6$, a 5$\sigma$ contrast of $\Delta H=7.3$ mag was achieved, 
and thus the CC was detected with $>5\sigma$.

The companion has a flux ratio of $0.00200 \pm 0.00033$, corresponding to $\Delta H=6.75 \pm 0.18$ mag. It is located at an angular separation of $0.605 \pm 0.026$ arcsec and a position angle of $151.8^{\circ} \pm 2.5^{\circ}$. If the companion were to be physically associated with \epichj, this corresponds to a projected physical separation of $409 \pm 21$ AU at the 676 pc distance of \epichj\ \citep{Gaia18}.

\begin{table}
	\centering
	\caption{Properties of the Candidate Companion to \epichj}
	\label{tab:companion}
	\begin{tabular}{lccc} 
		\hline
        Parameter & Unit & Value &  Uncertainty \\
        \hline
        \multicolumn{4}{l}{Measured Parameters} \\
        $\rho$ & arcsec & $0.605$ & $0.026$ \\
        PA & $^{\circ}$ & $151.8$ & $2.5$ \\
        $f_{\mathrm{CC}}/f_{A}$ & & $0.00200$ & $0.00033$ \\
        $\Delta H$ & mag & $6.75$ & $0.18$ \\
        \hline
        \multicolumn{4}{l}{Estimated Parameters} \\
        $a_{\perp}$ & AU & $409$ & $21$ \\
        $M_{\mathrm{CC}}$ & \msun & $0.145$ & $0.015$ \\
        Spec. Type & & M5-6V & \\
        \hline
        \multicolumn{4}{p{200pt}}{Notes: $f_{\mathrm{CC}}/f_{A}$ is the flux ratio of the companion with respect to the primary in the $H$ band. $a_{\perp}$ is the projected separation.}
	\end{tabular}
\end{table}

\begin{figure}
	\includegraphics[width=\columnwidth]{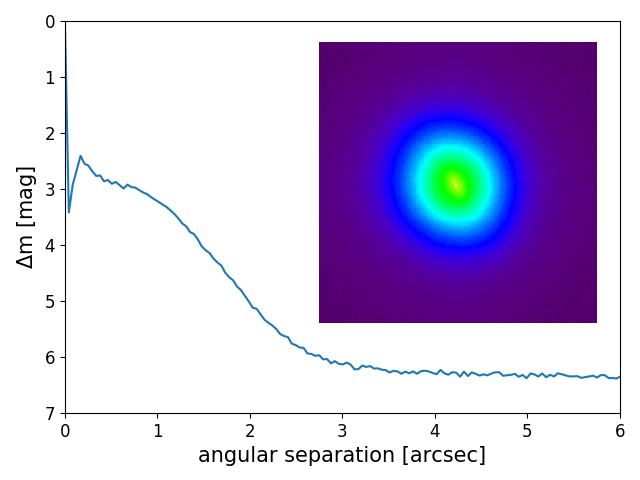}\\
    \caption{Clear filter contrast curve as a function of angular separation up to $6.0$ arcsec from \epicws\ obtained with the FastCam camera at TCS. The solid line indicates the $5\sigma$ detection limit for secondary sources, none of which are detected. The inset shows the $6\times 6$ arcsec combined image of \epicws. North is up and East is left.}
    \label{fig:FastCam}
\end{figure}

We obtained the observations of \epicws using the FastCam lucky imaging camera \citep{Oscoz08} on the 1.52 m Telescopio Carlos S\'anchez at Teide Observatory, Spain. FastCam is a very low noise and fast readout speed electron multiplying CCD camera with $512 \times 512$ pixels (with a physical pixel size of 16 $\mu$m, and a field of view of $21.2\times 21.2$ arcsec). We observed \epicws\ on 2018 March 20 UT, obtaining 5,000 individual frames with 50 ms exposure times through a clear filter. The conditions were clear with 1.7 arcsec seeing and we obtained a Strehl ratio of 0.06. In order to construct a high resolution, long-exposure image, each individual frame was bias-subtracted, aligned and co-added and then processed with the FastCam dedicated software 
\citep{Labadie10, Jodar13}.
Fig.~\ref{fig:FastCam} shows the contrast curve that was computed based on the scatter within the annulus as a function of angular separation from the target centroid. We used a high resolution image constructed by co-addition of the $30$ per cent best images, so that it had a 75~s total exposure time, which achieved a $5\sigma$ detection limit better than $\sim6$ magnitudes outside 3 arcsec separation, and better than $\sim3$ magnitudes outside $\sim50$ mas. We did not detect any secondary sources within these limits inside 6.0 arcsec of \epicws.

\subsection{Ground-based photometry}

In order to perform additional false positive vetting of \epichj, we obtained follow-up ground-based photometric observations of several transits. We obtained these observations with two different facilities. 

First, we observed with the DEMONEXT automated 0.51 m robotic telescope \citep{Villanueva18}, located at Winer Observatory, Arizona, USA. We observed the transits of 2017 November 14 and 22 UT in alternating Sloan $g$ and $i$ filters, using 20 s exposures in each band, and binned together each set of five consecutive exposures in each band in order to increase the signal-to-noise. We obtained these pseudo-simultaneous multi-band observations in order to verify that the transit is achromatic, excluding certain blended eclipsing binary scenarios. The observations were obtained through thin clouds, degrading the expected photometric performance, but nonetheless we detect the transits of \epichj~b. The transits in both $g$ and $i$ appear to be slightly shallower than that in the {\it K2} light curve; however, given the precision of the DEMONEXT data and systematic uncertainties in the depth associated with detrending the DEMONEXT data, we do not consider these differences to be significant. Furthermore, if these depth differences were due to dilution by another star, we would not expect {\it both} bands to exhibit smaller transit depths than the {\it K2} data unless the contaminating star were to be of a very similar \teff\ to \epichj. 
Although the DEMONEXT data are much less precise than the {\it K2} photometry, we nonetheless include them in our fits (\S\ref{sec:fitting}) as the longer time baseline improves the precision of our ephemeris measurement. We show these data in Fig.~\ref{fig:models1830}.

We additionally obtained observations of the partial transits of 2018 February 18 and March 10 UT using the Multicolor Simultaneous Camera for studying Atmospheres of Transiting exoplanets 2 \citep[MuSCAT2;][]{Narita18}, which is also on the Telescopio Carlos S\'anchez. 
MuSCAT2 observes simultaneously in the $g$, $r$, $i$, and $z$ bands, using a set of dichroics to split the light between four separate cameras. The transit depth in all four MuSCAT2 bands for both partial transits is consistent with that from {\it K2}, giving no evidence for a chromatic transit depth and indicating that the inconsistent transit depths in the DEMONEXT data are indeed likely due to systematics and problems with the detrending. We also include the MuSCAT2 data in our simultaneous fits. We also show these data in Fig.~\ref{fig:models1830}. 

The transit of \epicws~b is shallower than that of \epichj~b and not as easily detectable from the ground. We therefore did not obtain any follow-up transit photometry of \epicws~b.

\subsection{High-resolution spectroscopy}
\label{sec:specobs}

After identification of the transiting planet candidates around \epichj\ and \epicws, we obtained reconnaissance spectroscopy observations in order to measure the stellar parameters and rule out obvious false positive scenarios (such as eclipsing binaries with two sets of strong lines). We obtained these spectra with the 2.7 m Harlan J.\ Smith Telescope at McDonald Observatory, Texas, USA, and its Robert G.\ Tull coud\'e spectrograph \citep{Tull95}. We used the spectrograph in its TS23 configuration, which gives a spectral resolving power of $R=60,000$ over the range 3750-10200 \AA, with complete spectral coverage below 5691 \AA. We obtained a spectrum of each \epichj\ and \epicws\ with a signal-to-noise ratio (SNR) per-resolution element of 35 and 50 at 5650 \AA\ for \epichj\ and \epicws, respectively. We measured stellar parameters from these spectra using the Kea spectral analysis tool \citep{EndlCochran16}. 

The reconnaissance spectra indicated that both \epichj\ and \epicws\ were good targets for precise radial velocity (RV) follow-up in order to confirm these planet candidates and measure their masses: both stars are relatively slowly rotating and have high surface gravity, and neither of them show any evidence of multiple lines in the spectra. We obtained precise RVs with three different facilities; these have higher SNRs and so superseded the reconnaissance spectra in our further analyses.

We obtained observations of both \epichj\ and \epicws\ using the FIbre-fed \'Echelle Spectrograph \citep[FIES;][]{FrandsenLindberg99,Telting14} on the 2.56 m Nordic Optical Telescope (NOT) at the Observatorio del Roque de los Muchachos, La Palma, Spain. We used FIES' {\it high-res} mode, which provides spectra with $R=67,000$ over the range 3600-9000 \AA. We used the same observing strategy as in \cite{Gandolfi15}; at each observation epoch we obtained three consecutive exposures to facilitate cosmic ray removal, bracketed by $120-180$-second ThAr exposures in order to trace any drift of the spectrograph. We obtained eighteen RVs of \epichj\ with FIES between 2017 November 16 and 2018 March 11 UT, and twelve of \epicws\ between 2017 December 26 and 2018 February 15 UT, as part of the observing programs 56-010, 56-112, and 56-209.

We also obtained observations of both targets using the High Accuracy Radial velocity Planet Searcher for the Northern hemisphere \citep[HARPS-N;][]{Cosentino12} spectrograph on the 3.58 m Telescopio Nazionale Galileo (TNG), also at La Palma. HARPS-N is a fibre-fed cross-dispersed \'echelle spectrograph, and provides spectra with $R=115,000$ over the range 3830-6900 \AA. We observed \epicws\ nine times between 2017 December 27 and 2018 February 22 UT, and \epichj\ twice on 2018 February 19 UT, as part of the programs A36TAC\_12, CAT17B\_99, and OPT17B\_059. We did not utilize the two HARPS-N RVs for \epichj\ in our fits as their poor phase coverage results in little constraining power on the fits. Additionally, one HARPS-N RV of \epicws\ was excluded from the fit due to poor signal-to-noise. 

Finally, we obtained observations of \epicws\ using the High Accuracy Radial velocity Planet Searcher \citep[HARPS;][]{Mayor03}. The spectrograph is essentially identical to HARPS-N, and is located on the ESO 3.6 m telescope at La Silla Observatory, Chile. We obtained ten RV observations of \epicws\ using HARPS between 2018 January 25 and May 13 UT under the programs 0100.C.0808 and 0101.C-0829. One HARPS RV was excluded from our fits due to an incorrect flux correction by the HARPS pipeline.

For both HARPS and HARPS-N we reduced the data using the dedicated off-line pipelines and extracted the RVs via cross-correlation with a G2 numerical mask \citep{Baranne96, Pepe02}. For FIES we reduced the data as described in \cite{Gandolfi15}, using standard IRAF and IDL routines, which include bias subtraction, flat fielding, order tracing and extraction, and wavelength calibration. The FIES RVs of \epichj\ were derived by cross-correlating the observed spectra against a high SNR (>300) spectrum of the RV standard star HD\,50692 observed with the same instrument set-up as \epichj. For \epicws, relative RVs were extracted by cross-correlating the observed FIES data against the first stellar spectrum.

\begin{figure*}
    \includegraphics[width=\textwidth]{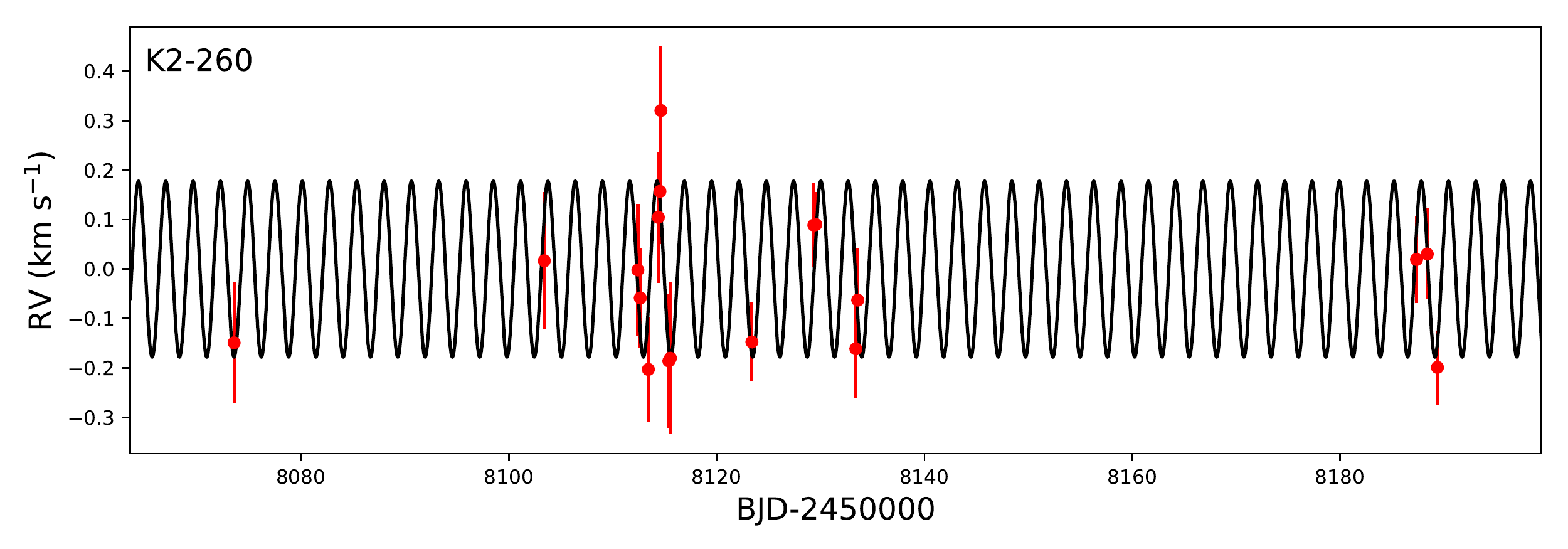}\\
	\includegraphics[width=\textwidth]{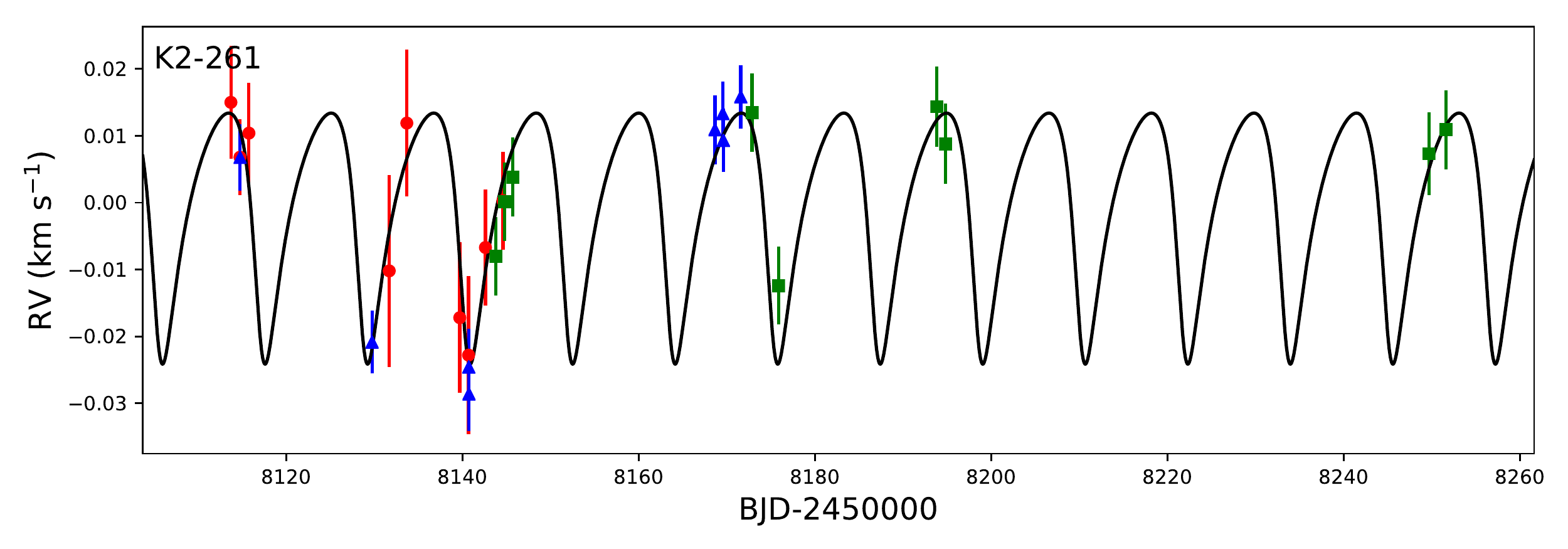}\\
    \caption{Radial velocity measurements of \epichj\ (top) and \epicws\ (bottom), along with the best-fit orbit models computed in \S\ref{sec:fitting}. Data from FIES are shown as circles (red in the online version), HARPS-N as triangles (blue online), and HARPS as squares (green online). The error bars incorporate both the internal RV errors and the best-fit RV jitter.}
    \label{fig:allRVs}
\end{figure*}

We show these data in Fig.~\ref{fig:allRVs}, and list all of our RV measurements  in Table~\ref{tab:RVs}, along with the cross-correlation function bisector span (BIS) and full width at half maximum (FWHM), and the Ca\,{\sc ii} H\,\&\,K chromospheric activity index $\log R'_\mathrm{HK}$ (the latter measured only from the HARPS-N and HARPS spectra). We do not find any significant correlation between the RV measurements and the BIS, as well as between the RVs and the FWHMs, for either target, and we robustly detect the stellar RV variations induced by both transiting objects, confirming both \epichj~b and \epicws~b as {\it bona fide} planets.

\begin{table*}
	\centering
	\caption{Radial velocity measurements of \epichj\ and \epicws}
	\label{tab:RVs}
	\begin{tabular}{lrrrrrrlrr} 
		\hline
		BJD$_{\mathrm{TDB}}$ & RV & $\sigma_{\mathrm{RV}}$ & BIS & FWHM & $\log R'_\mathrm{HK}$ & $\sigma_{\log R'_\mathrm{HK}}$ & $T_{\mathrm{exp}}$ & SNR & Instrument \\
        	& (\kms) & (\kms) & (\kms) & (\kms) & & & (s) & \\
		\hline
        \epichj \\
$2458073.56279$ & $28.9213$ & $0.1196$ & $0.4946$ & $28.3623$ & & & $3600$ & $23$ & FIES \\ 
$2458103.41726$ & $29.0872$ & $0.1358$ & $0.5607$ & $28.0208$ & & & $3600$ & $25$ & FIES \\ 
$2458112.41796$ & $29.0684$ & $0.1306$ & $0.7535$ & $28.5843$ & & & $3600$ & $21$ & FIES \\ 
$2458112.64764$ & $29.0119$ & $0.0966$ & $0.2418$ & $28.1836$ & & & $3600$ & $28$ & FIES \\ 
$2458113.42787$ & $28.8675$ & $0.1017$ & $0.4048$ & $28.2302$ & & & $3600$ & $28$ & FIES \\ 
$2458114.38939$ & $29.1748$ & $0.1296$ & $0.7658$ & $28.3094$ & & & $3600$ & $20$ & FIES \\ 
$2458114.54165$ & $29.2274$ & $0.1025$ & $0.6633$ & $28.4909$ & & & $3600$ & $24$ & FIES \\ 
$2458114.63722$ & $29.3908$ & $0.1274$ & $0.5007$ & $28.2135$ & & & $3600$ & $18$ & FIES \\ 
$2458115.40316$ & $28.8845$ & $0.1320$ & $0.6163$ & $28.4217$ & & & $3600$ & $21$ & FIES \\ 
$2458115.55094$ & $28.8900$ & $0.1512$ & $0.6702$ & $28.1998$ & & & $3600$ & $14$ & FIES \\ 
$2458123.39303$ & $28.9230$ & $0.0753$ & $0.2370$ & $28.1771$ & & & $3600$ & $29$ & FIES \\ 
$2458129.35492$ & $29.1592$ & $0.0805$ & $0.2508$ & $29.0288$ & & & $3600$ & $27$ & FIES \\ 
$2458129.53697$ & $29.1604$ & $0.0599$ & $0.2017$ & $28.1717$ & & & $3600$ & $26$ & FIES \\ 
$2458133.39617$ & $28.9090$ & $0.0948$ & $0.3268$ & $28.5724$ & & & $3600$ & $28$ & FIES \\ 
$2458133.57782$ & $29.0075$ & $0.1015$ & $0.2893$ & $28.2565$ & & & $3600$ & $26$ & FIES \\ 
$2458187.35731$ & $29.0897$ & $0.0843$ & $0.3350$ & $29.2858$ & & & $3600$ & $25$ & FIES \\ 
$2458188.38929$ & $29.1007$ & $0.0879$ & $0.7315$ & $28.1793$ & & & $3600$ & $26$ & FIES \\ 
$2458189.36976$ & $28.8715$ & $0.0699$ & $0.1347$ & $28.8898$ & & & $3600$ & $24$ & FIES \\ 
$2458169.40257$   &   $29.279$    &     $0.052$ & $-1.206$ & $26.5556$ & $-4.594$ & $0.062$ & $3600$ & $18$ & HARPS-N \\
$2458169.44866$   &   $29.366$    &     $0.050$ & $-6.231$ & $27.0874$ & $-4.674$ & $0.071$ & $3600$ & $19$ & HARPS-N \\

        \hline
        \epicws \\
$2458113.71551$ & $ 0.0000$ & $0.0079$ & $ 0.0005$ & $12.4310$ & & & $3600$ & $52$ & FIES \\ 
$2458114.74539$ & $-0.0052$ & $0.0049$ & $ 0.0089$ & $12.3931$ & & & $3600$ & $57$ & FIES \\ 
$2458115.75979$ & $-0.0045$ & $0.0068$ & $ 0.0065$ & $12.4364$ & & & $3600$ & $63$ & FIES \\ 
$2458131.68086$ & $-0.0300$ & $0.0140$ & $ 0.0130$ & $12.4362$ & & & $3600$ & $29$ & FIES \\ 
$2458133.67474$ & $-0.0068$ & $0.0095$ & $ 0.0081$ & $12.4448$ & & & $3600$ & $38$ & FIES \\ 
$2458139.68280$ & $-0.0340$ & $0.0110$ & $ 0.0360$ & $12.4553$ & & & $3600$ & $37$ & FIES \\ 
$2458140.68126$ & $-0.0350$ & $0.0120$ & $ 0.0130$ & $12.3920$ & & & $3600$ & $38$ & FIES \\ 
$2458142.59534$ & $-0.0284$ & $0.0084$ & $ 0.0139$ & $12.4582$ & & & $3600$ & $43$ & FIES \\ 
$2458144.59629$ & $-0.0115$ & $0.0068$ & $ 0.0082$ & $12.3859$ & & & $3600$ & $44$ & FIES \\ 
$2458146.69263$ & $ 0.0020$ & $0.0110$ & $ 0.0200$ & $12.4276$ & & & $3600$ & $35$ & FIES \\ 
$2458163.53606$ & $-0.0179$ & $0.0056$ & $ 0.0165$ & $12.4710$ & & & $3600$ & $56$ & FIES \\ 
$2458164.59738$ & $-0.0369$ & $0.0050$ & $-0.0041$ & $12.4503$ & & & $3600$ & $55$ & FIES \\ 
$2458114.75801$ & $3.3424$ & $0.0019$ & $-0.0058$ & $7.4426$ & $-5.139$ & $0.043$ & $900$ & $43$ & HARPS-N \\ 
$2458129.73171$ & $3.3148$ & $0.0011$ & $-0.0087$ & $7.4581$ & $-5.129$ & $0.019$ & $1800$ & $68$ & HARPS-N \\ 
$2458140.70807$ & $3.3110$ & $0.0034$ & $-0.0133$ & $7.4584$ & $-5.204$ & $0.163$ & $1200$ & $29$ & HARPS-N \\ 
$2458140.72396$ & $3.3070$ & $0.0030$ & $ 0.0024$ & $7.4599$ & $-4.943$ & $0.091$ & $1200$ & $33$ & HARPS-N \\ 
$2458168.63918$ & $3.3465$ & $0.0025$ & $-0.0006$ & $7.4408$ & $-5.169$ & $0.066$ & $2100$ & $36$ & HARPS-N \\ 
$2458169.52666$ & $3.3489$ & $0.0014$ & $-0.0066$ & $7.4393$ & $-5.077$ & $0.026$ & $3300$ & $55$ & HARPS-N \\ 
$2458169.59477$ & $3.3449$ & $0.0013$ & $-0.0069$ & $7.4464$ & $-5.056$ & $0.021$ & $3300$ & $60$ & HARPS-N \\ 
$2458169.70060^{\dagger}$ & $3.2810^{\dagger}$ & $0.0440^{\dagger}$ & $-0.1753^{\dagger}$ & $7.2957^{\dagger}$ & $-4.545^{\dagger}$ & $0.451^{\dagger}$ & $267^{\dagger}$ & $4^{\dagger}$ & HARPS-N$^{\dagger}$ \\ 
$2458171.56866$ & $3.3514$ & $0.0011$ & $-0.0065$ & $7.4418$ & $-5.111$ & $0.017$ & $1500$ & $70$ & HARPS-N \\ 

$2458143.78984$ & $3.3331$ & $0.0013$ & $-0.0093$ & $7.4954$ & $-5.177$ & $0.031$ & $1500$ & $64$ & HARPS \\ 
$2458144.77455$ & $3.3412$ & $0.0014$ & $-0.0054$ & $7.4928$ & $-5.207$ & $0.039$ & $1200$ & $58$ & HARPS \\ 
$2458145.71595$ & $3.3449$ & $0.0014$ & $-0.0050$ & $7.4817$ & $-5.208$ & $0.040$ & $1500$ & $57$ & HARPS \\ 
$2458171.71521^{\dagger}$ & $3.3438^{\dagger}$ & $0.0016^{\dagger}$ & $-0.0179^{\dagger}$ & $7.6649^{\dagger}$ & & & $1500^{\dagger}$ & $71^{\dagger}$ & HARPS$^{\dagger}$ \\ 
$2458172.84902$ & $3.3546$ & $0.0012$ & $-0.0030$ & $7.4881$ & $-5.255$ & $0.043$ & $1500$ & $67$ & HARPS \\ 
$2458175.84594$ & $3.3287$ & $0.0011$ & $-0.0056$ & $7.4916$ & $-5.184$ & $0.035$ & $1500$ & $76$ & HARPS \\ 
$2458193.81413$ & $3.3555$ & $0.0018$ & $-0.0038$ & $7.4958$ & $-5.081$ & $0.046$ & $1200$ & $46$ & HARPS \\ 
$2458194.80916$ & $3.3499$ & $0.0018$ & $0.0030$ & $7.4968$ & $-5.010$ & $0.040$ & $1200$ & $47$ & HARPS \\ 
$2458249.65186$  & $3.3484$ &  $0.0023$ & $0.0053$ & $7.5094$ & $-5.303$ &   $0.099$ & $1200$ & $37$ & HARPS \\
$2458251.59291$ & $3.3520$ & $0.0016$ & $-0.0014$ & $7.4957$ & $-5.194$ & $0.048$ & $1200$ & $52$ & HARPS \\
        \hline
        \multicolumn{10}{p{450pt}}{Notes: We do not use the two HARPS-N spectra of \epichj\ in our fits as the phase coverage is insufficient to provide good constraints upon the fits, and also excluded the two spectra marked as $^{\dagger}$ from our fits due to low signal-to-noise and other problems. The FIES RVs of \epicws\ are measured with respect to the first spectrum, the RV of which is arbitrarily set to zero. See Table~\ref{tab:planetpars} for the best-fit RV offsets for each dataset. We did not measure the activity index $\log R'_\mathrm{HK}$ from our FIES spectra as our pipeline is not configured to do so. The bisector spans quoted for the HARPS-N data on \epichj\ are likely inaccurate as they were derived using the pipeline G2 CCF mask, which is not a good match to the F6 spectral type of the target. The quoted signal-to-noise ratio (SNR) is the per-pixel SNR measured at 5500 \AA. }
	\end{tabular}
\end{table*}

\section{Analysis}

\subsection{Spectral analysis and stellar parameters}
\label{sec:specanalysis}

In order to derive more precise stellar parameters than were found using our reconnaissance spectra (\S\ref{sec:specobs}), we analysed our co-added HARPS-N spectra of \epicws, and our co-added FIES spectra of \epichj. The spectroscopic analysis was performed within the \textsc{iSpec} framework \citep{BlancoCuaresma14}. We derived spectroscopic parameters ($T_{\text{eff}}$, $\log g$, \feh, $v_{\text{mic}}$) by fitting synthetic spectra computed using \textsc{moog} \citep{Sneden73} and ATLAS9 model atmospheres \citep{CastelliKurucz04} to the co-added spectra. We computed the broadening function by convolving the observed spectrum with a sharp-lined template. The broadening function was fitted with a broadened line profile to obtain \vsinistar\ and the macroturbulent velocity $v_{\text{mac}}$. For \epichj\ we fixed the microturbulent velocity $v_{\text{mic}}$ using the values from \cite{Doyle14} expected based upon the \teff, \loggstar\, and \feh\ in the fit. 

For \epichj\ we could not obtain a good measurement of \loggstar\ from the spectra due to insufficient signal-to-noise and the moderate rotational broadening (\vsinistar$=16.0 \pm 2.0$ \kms). We therefore estimated $\loggstar=4.15_{-0.04}^{+0.02}$ using the mean stellar density derived from the circular-orbit transit fit (\S\ref{sec:fitting}) and \textsc{basta} isochrones \citep{SilvaAguirre15} with a flat prior over the range $5500$ K$<$\teff$<7500$ K. We fixed \loggstar\ to this value in the spectral analysis as the uncertainties are much smaller than generally obtainable from spectroscopy alone. For both stars we leveraged the precise {\it Gaia} DR2 parallaxes \citep{Gaia16b,Gaia18} to fit the stellar SEDs, using the $BVgriJHK$-band photometry listed in Table~\ref{tab:litpars}. For \epichj\ we included reddening of $E(B-V)=0.18 \pm 0.006$ from \cite{Green18}, while for \epicws\ we assumed $E(B-V)=0$. 
The stellar parameters for the two stars are listed in Table~\ref{tab:starpars}.

\begin{figure}
	\includegraphics[width=1.0\columnwidth]{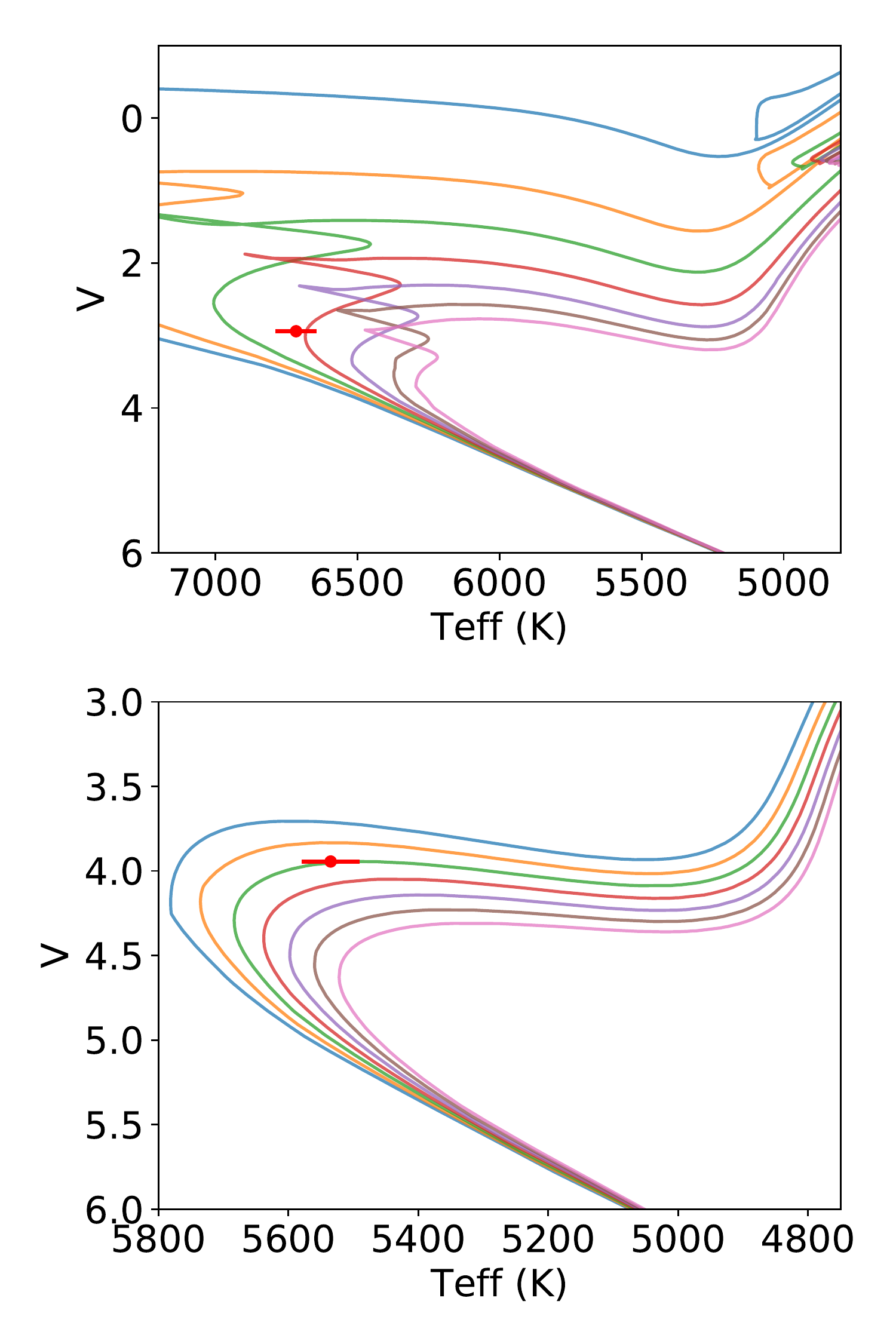}\\
    \caption{H-R diagrams showing \epichj\ (top) and \epicws\ (bottom). For \epichj\ the lines show BaSTI isochrones of, from top to bottom, 0.5 (coloured blue online), 1.0 (orange), 1.5 (green), 2.0 (red), 2.5 (purple), 3.0 (brown), and 3.5 (pink) Gyr for \feh$=-0.1$, while for \epicws the lines show isochrones with ages of, from left to right, 7 (blue), 8 (orange), 9 (green), 10 (red), 11 (purple), 12 (brown), and 13 (pink) Gyr for \feh$=+0.4$. Coincidentally, both \epichj\ and \epicws\ are located near the main sequence turn-off, although \epicws\ has already passed this point whereas \epichj\ is only now reaching it.}
    \label{fig:HRdiag8078}
\end{figure}

\begin{figure}
	\includegraphics[width=1.1\columnwidth, trim={0 3cm 0 4cm}, clip]{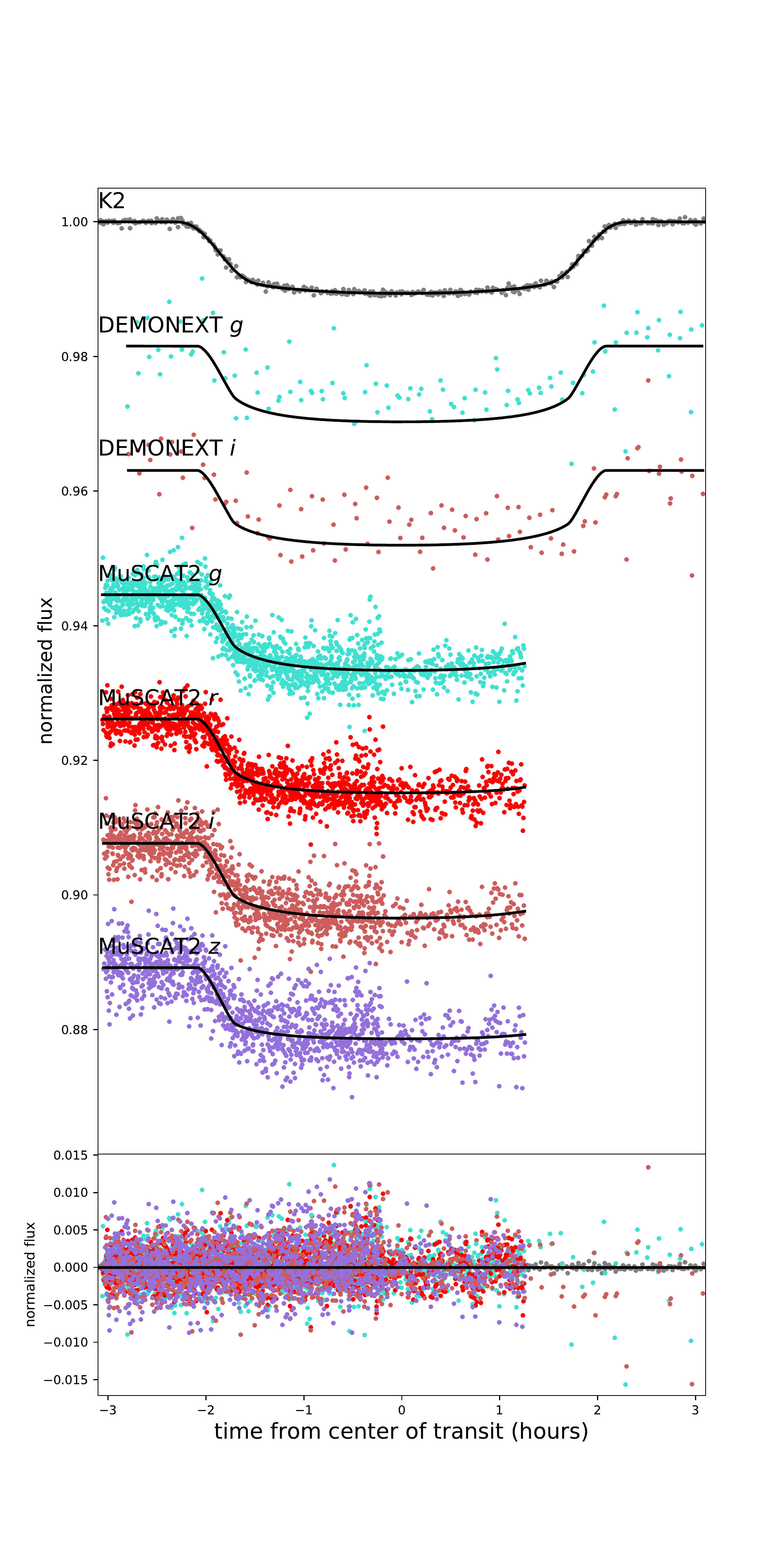}\\
    \vspace{-24pt}
    \includegraphics[width=1.1\columnwidth, trim={0 0 0 1.25cm}, clip]{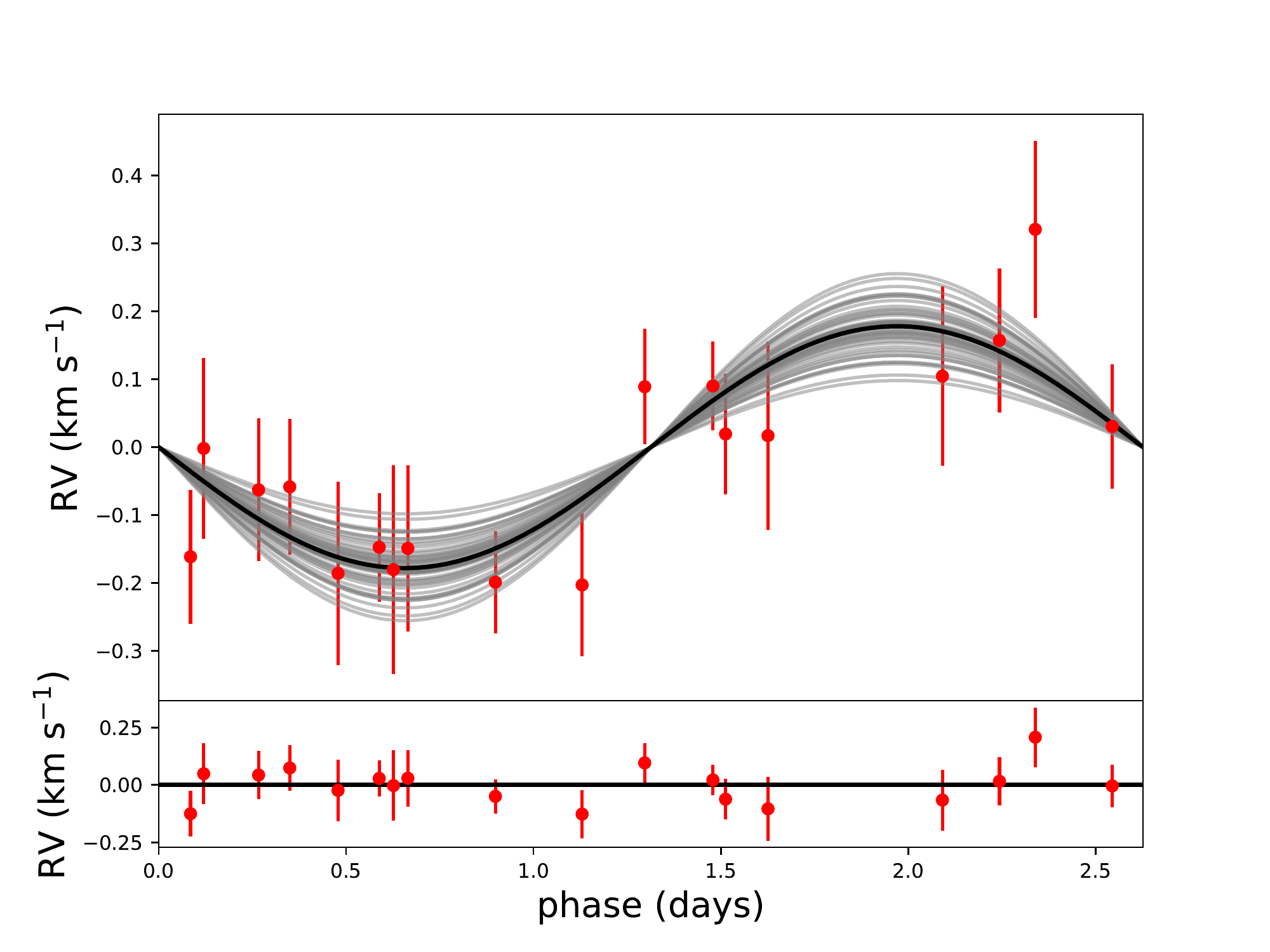}\\
    \caption{Top: light curves of \epichj, phase-folded on the transit period and with the best-fit transit models overplotted. 
    The transits in each band-instrument combination are offset vertically by an arbitrary amount for clarity; note that the DEMONEXT and MuSCAT2 light curves both contain two individual transits overplotted. 
    Bottom: RV measurements for \epichj\ phased to the transit period, with the best-fit model from the circular orbital fit overplotted in black. The RV curves corresponding to 50 random draws from the posterior distributions are overplotted in grey. The error bars incorporate both the internal RV errors and the best-fit RV jitter.}
    \label{fig:models1830}
\end{figure}

We inferred the stellar masses, radii, and ages using the \textsc{basta} code to fit the measured \teff, \feh, and mean stellar density to an isochrone grid from the BaSTI database\footnote{\url{http://ia2.oats.inaf.it/archives/basti-a-bag-for-stellar-tracks-and-isochrones}} computed using the \textsc{franec} stellar evolutionary models \citep{Pietrinferni04}. Our derived parameters are also listed in Table~\ref{tab:starpars}. Thanks to the location of both stars near the main sequence turn-off we are able to measure relatively precise ages of $1.9 \pm 0.3$ Gyr for \epichj\ and $8.8_{-0.3}^{+0.4}$ Gyr for \epicws. We show the position of both stars in the H-R diagram in Fig.~\ref{fig:HRdiag8078}.

We estimated spectral types for the stars based upon the \teff-spectral type calibration of \cite{PecautMamajek13}, obtaining spectral types of F6 and G7, respectively. \epicws\ lies just past the main sequence turn-off, and so we quote a spectral type of G7IV/V, while \epichj\ lies at the main sequence turn-off, resulting in a spectral type of F6V. 

\begin{table}
	\centering
	\caption{Stellar parameters of \epichj\ and \epicws}
	\label{tab:starpars}
	\begin{tabular}{lcc} 
		\hline
		Parameter & \epichj & \epicws  \\
		\hline
        Measured Parameters \\
		\teff\ (K) & $6367 \pm 250$ & $5537 \pm 71$ \\
        \loggstar (cgs) & $4.15$ (fixed) & $4.21 \pm 0.11$ \\
        \feh\ (dex) & $-0.14 \pm 0.15$ & $0.36 \pm 0.06$ \\
        \vsinistar\ (\kms) & $16.0 \pm 2.0$ & $2.8 \pm 0.5$ \\
        $v_{\mathrm{mic}}$ (\kms) & 1.6 (fixed) & $1.3 \pm 0.1$ \\
        $v_{\mathrm{mac}}$ (\kms) & $10.0 \pm 2.0$ & $2.2 \pm 0.5$ \\

		\hline
        Derived Parameters \\
        Spec. Type & F6V & G7IV/V \\
        \mstar\ (\msun) & $1.39_{-0.06}^{+0.05}$ & $1.10_{-0.02}^{+0.01}$ \\
        \rstar\ (\rsun) & $1.69 \pm 0.03$ & $1.65 \pm 0.04$ \\
        $\rho_{\star}$ ($\rho_{\odot}$) & $0.2841 \pm 0.0071$  & $0.248 \pm 0.021$ \\ 
        age (Gyr) & $1.9 \pm 0.3$ & $8.8_{-0.3}^{+0.4}$ \\
        
        \hline
        
	\end{tabular}
\end{table}

\subsection{$UVW$ space motion}

We calculated the $UVW$ space motion for both \epichj\ and \epicws, since these stars have {\it Gaia} DR2 6-d kinematic measurements \citep[i.e., position, parallax, proper motion, and radial velocity;][]{Gaia16b,Gaia18}. We used the IDL routine \textsc{gal\_uvw}\footnote{\url{https://idlastro.gsfc.nasa.gov/ftp/pro/astro/gal_uvw.pro}}, which is based upon \cite{JohnsonSoderblom87}. Using the local standard of rest (LSR) definition from \cite{Coskunoglu11}, we found that \epichj\ has a space velocity of $(U,V,W)=(15.8 \pm 2.6, -6.15 \pm 0.57, -9.46 \pm 0.73)$ \kms, while \epicws\ has $(U,V,W)=(-8.99 \pm 0.13, -29.14 \pm 0.49, -21.81 \pm 0.50)$ \kms, both in the LSR frame. We then used the methodology of \cite{Reddy06} to estimate the probability that these stars belong to various Galactic components. We find that \epichj\ has a 99 per cent probability of belonging to the Galactic thin disk, 1 per cent that it is a member of the thick disk, and $2\times10^{-3}$ per cent that it belongs to the halo. This is expected as F stars are relatively young. \epicws has a 97 per cent probability of belonging to the thin disk, 3 per cent that it is a member of the thick disk, and $8\times10^{-3}$ per cent that it is a member of the halo. The high \feh\ of \epicws\ found in \S\ref{sec:specanalysis} (\feh$=0.36 \pm 0.06$) is also consistent with its membership in the Galactic thin disk.

\subsection{Joint analysis of photometry and radial velocities}
\label{sec:fitting}

In order to measure the parameters of \epichj~b and \epicws~b, we simultaneously fit all of the available time-series photometry and RVs for each system. We use the same Python-based fitting code used by \cite{DT3}, but modified to allow for non-zero orbital eccentricity. It samples the likelihood function of model fits to the data using an affine-invariant Markov chain Monte Carlo (MCMC) algorithm as implemented in \textsc{emcee} \citep{ForemanMackey13}. 

We generated photometric transit models using the \textsc{batman} package \citep{Kreidberg15}. We assumed a quadratic limb darkening law, and used the triangular sampling method of \cite{Kipping13} to uniformly sample the relevant parameter space. We calculated expected limb darkening values in each band using the \textsc{jktld} code \citep{Southworth15} to interpolate limb darkening values to the best-fit stellar \teff\ and \loggstar\ measured earlier, from the tabulations provided for the \kepler\ bandpass by \cite{Sing10} and for the $g$, $r$, $i$, and $z$ bands by \cite{Claret04}. 

For circular orbital fits we simply modelled the RV curve as a cosine function, while for eccentric orbits we computed the RV model using the \textsc{RadVel} package \citep[][but note that we only used \textsc{RadVel} to compute the model and did not use the fitting functions provided by that package]{Fulton18}. We fit for an individual RV offset $\gamma$ for each facility, and also included an RV jitter term for each facility added in quadrature to the uncertainty on each RV data point.

All in all, our MCMCs included the following parameters: orbital period $P$, transit epoch $T_0$, radius ratio $R_P/R_{\star}$, impact parameter $b$, and the scaled semi-major axis $a/R_{\star}$; two quadratic limb-darkening parameters per photometric bandpass; the RV semi-amplitude $K$ and one RV offset $\gamma$ and one RV jitter term per RV facility; for the fits including an RV trend, a linear acceleration term $\dot{\gamma}$; and, in the case of eccentric orbital fits, $e\sin\omega$ and $e\cos\omega$. We set Gaussian priors upon each limb darkening value with a $\sigma$-width of 0.1, and for the other parameters used uniform priors within physically-allowed bounds, except as noted below. We performed fits with both circular and eccentric orbits for both planets, and we also performed fits including a linear RV trend for both systems in order to search for any evidence of RV trends due to additional objects in these systems. 

\subsubsection{\epichj}

For \epichj\ we do not find any evidence for an RV trend (i.e., $\dot{\gamma}$ is zero to within $1\sigma$), and set $3\sigma$ limits of $-2.7<\dot{\gamma}<2.7$ \ms\ day$^{-1}$. Our final fits for \epichj\ thus assumed $\dot{\gamma}=0$.
 For the eccentric fit for \epichj~b, we set a Gaussian prior upon $e\cos\omega$ with a central value of -0.0049 and a width $\sigma$ of 0.0048, based upon the best-fit value of this parameter from the analysis of the secondary eclipse (\S\ref{sec:2ndaryeclipse}).

We show the best-fit light curve and RV models for \epichj, along with the data, in Fig.~\ref{fig:models1830}. The best-fit eccentricity from the eccentric model is non-zero at a significance of only $1.5\sigma$, while $e\sin\omega$ is consistent with zero. Additionally, we separately measured that $e\cos\omega$ is also consistent with zero (\S\ref{sec:2ndaryeclipse}). Finally, the Bayesian information criterion (BIC) shows a strong preference for the circular model, with $\Delta$ BIC$=17.5$ in preference of the circular model. We therefore conclude that there is no compelling evidence that the orbit of \epichj~b is eccentric and adopt the circular fit. Nonetheless, in the interests of completeness we list the best-fit parameters from both fits in Table~\ref{tab:planetpars}.

\subsubsection{\epicws}

\begin{figure}
	\includegraphics[width=1.1\columnwidth, trim={0 0 0 1.25cm}, clip]{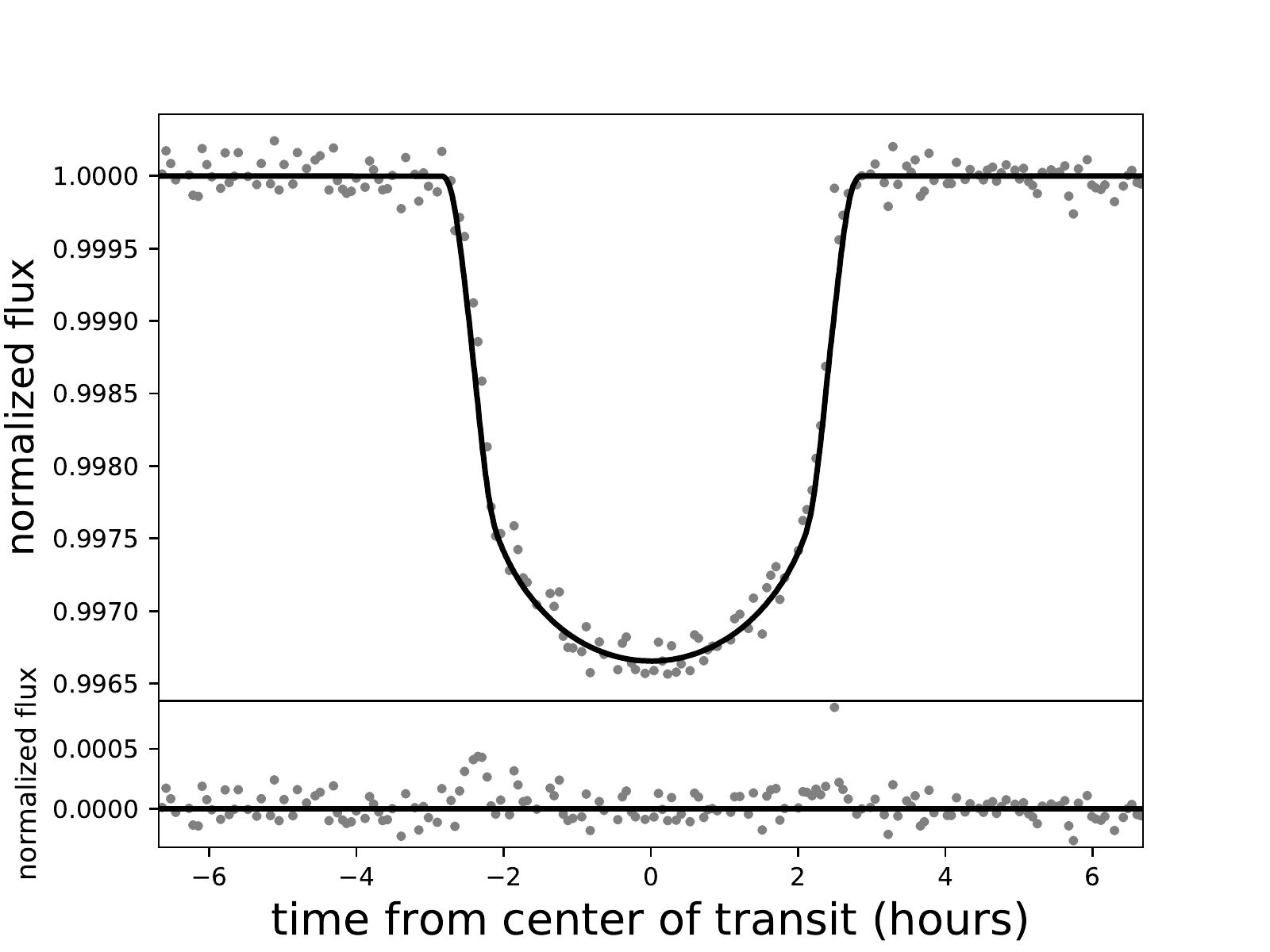}\\
    \includegraphics[width=1.1\columnwidth, trim={0 0 0 1.25cm}, clip]{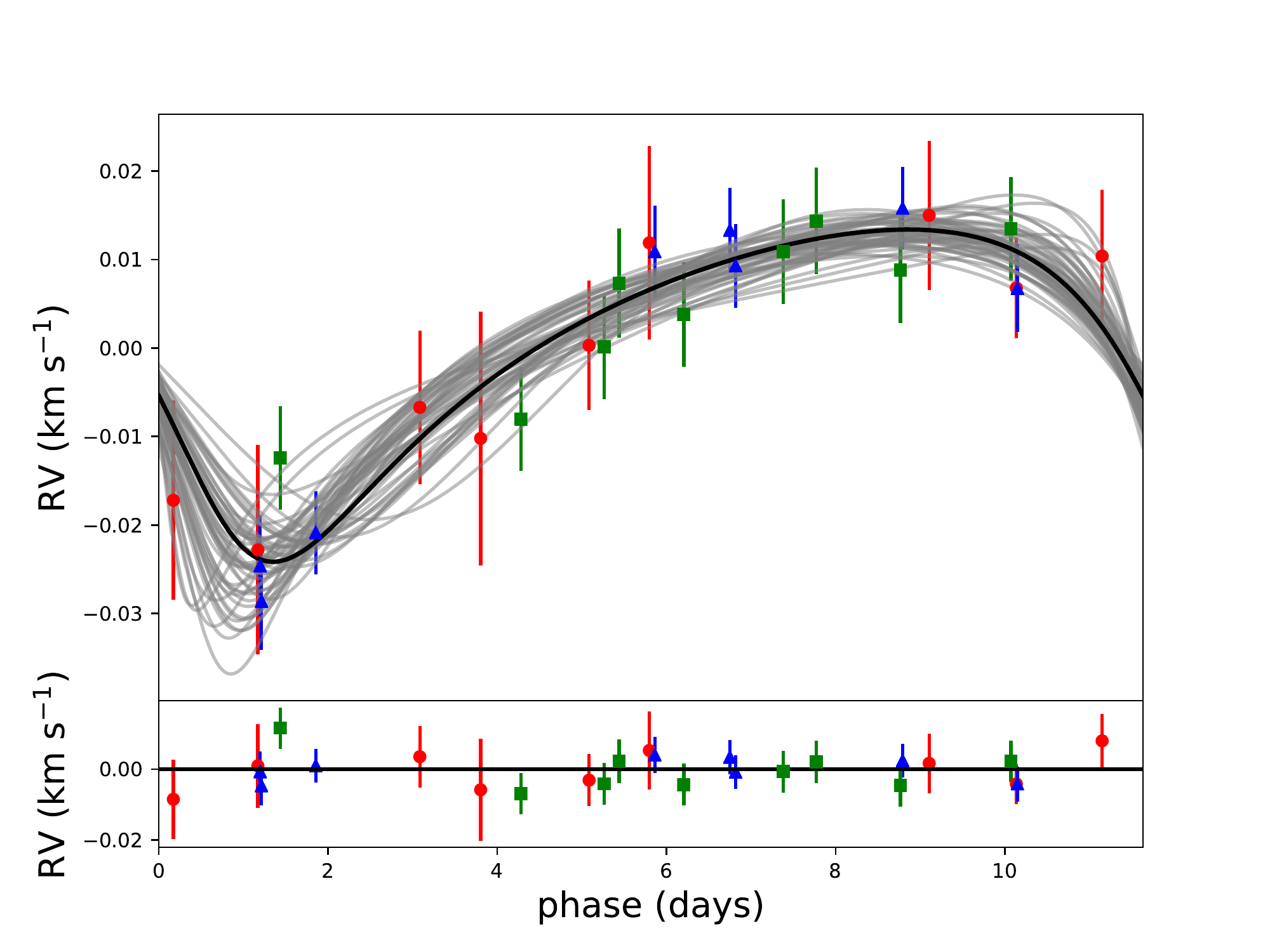}\\
    \vspace{-20pt}
    \caption{Top: {\it K2} light curve of \epicws, phase-folded on the transit period and with the best-fit transit model overplotted. 
    Bottom: RV measurements of \epicws\ from FIES (circles, coloured red online), HARPS-N (triangles, blue online), and HARPS (squares, green online), phased to the transit period and with the best-fit RV model overplotted in black, along with the models corresponding to 50 random draws from the posterior distributions in grey. The error bars incorporate both the internal RV errors and the best-fit RV jitter.} 
    \label{fig:models8078}
\end{figure}

We show the best-fit light curve model and {\it K2} data, and the radial velocity curve for the eccentric orbital fit, for \epicws\ in Fig.~\ref{fig:models8078}. For \epicws~b the eccentric fit prefers a eccentric orbit, with $e=0.39 \pm 0.15$. While this is formally only a $2.6\sigma$ detection of a non-zero eccentricity, the BIC prefers the eccentric model, with a $\Delta$ BIC value of 9.3 in preference of the eccentric model. Since a $\Delta$ BIC value of $<10$ is not conclusive, we also conducted a more fully Bayesian model comparison. We computed the Bayes factor for the model comparison, using the marginal likelihood approximation of \cite{ChibJeliazkov01}; we ported the implementation of this approximation from the \textsc{BayesianTools} R library \citep{BayesianTools} to Python for use with our MCMC chains. Doing so, we estimate a Bayes factor of 237 in preference of the eccentric model, indicating a decisive preference for the presence of orbital eccentricity for \epicws.

Including an RV trend results in a best-fit value of $\dot{\gamma}=0.106_{-0.057}^{+0.051}$ \ms\ d$^{-1}$, which is non-zero at the $2.1\sigma$ level and is thus marginally significant. This also improves the fit to the RV data. Including the trend, however, requires a more eccentric planetary orbit ($e=0.52^{+0.12}_{-0.13}$), which results in a poorer fit to the {\it K2} light curve as the eccentricity affects the detailed light curve shape. Furthermore, this requires an implausibly low mean stellar density of $0.032^{+0.061}_{-0.021}\,\rho_{\odot}$, which is $3.3\sigma$ discrepant with the value expected from the spectroscopic stellar parameters (while the value from the eccentric fit with no trend is only $0.9\sigma$ from this expected value, and the circular fit is $3.0\sigma$ away). While the model including the trend is disfavoured by only $\mathrm{\Delta}$ BIC$\mathrm={0.3}$ with respect to the eccentric model, we disfavour this model based upon the implausibly low $\rho_{\star}$ and we conclude that the inclusion of an RV slope results in a physically disallowed solution. For the above reasons we therefore adopt the eccentric fit with no RV trend as our preferred solution for \epicws; 
most of the planetary parameters are consistent between the circular and eccentric fits, although the RV semi-amplitude $K$ and thus the planetary mass is slightly smaller for the circular fit. We list the parameters from both of these fits with $\dot{\gamma}$ fixed to zero in Table~\ref{tab:planetpars}.

\begin{table*}
	\centering
	\caption{Planetary parameters of \epichj~b and \epicws~b}
	\label{tab:planetpars}
	\begin{tabular}{lcccc} 
		\hline
		Parameter (unit) & \epichj~b & \epichj~b & \epicws~b & \epicws~b \\ 
         & circular fit & eccentric fit & circular fit & eccentric fit \\ 
         & (preferred) & & & (preferred) \\
		\hline
 $\Delta$ BIC (circular $-$ eccentric) & \multicolumn{2}{c}{$-17.5$} & \multicolumn{2}{c}{$9.3$} \\
        \hline
        \multicolumn{5}{l}{MCMC Parameters} \\

$P$ (days)  &  $2.6266657 \pm 0.0000018$ &  $2.6266657^{+0.0000020}_{-0.0000018}$ &  $11.63344 \pm 0.00012$ &  $11.63344 \pm 0.00012$ \\ 
$T_0$ (BJD)  &  $2457820.738135 \pm 0.000090$ &  $2457820.738133^{+0.000091}_{-0.000098}$ &  $2457906.84115 \pm 0.00045$ &  $2457906.84084^{+0.00054}_{-0.00067}$ \\ 
$R_P/R_{\star}$  &  $0.09731^{+0.00032}_{-0.00025}$ &  $0.09731^{+0.00033}_{-0.00027}$ &  $0.05281^{+0.00104}_{-0.00051}$ &  $0.05293^{+0.00096}_{-0.00051}$ \\ 
$a/R_{\star}$  &  $5.291^{+0.033}_{-0.073}$ &  $5.34^{+0.92}_{-1.00}$ &  $17.79^{+0.57}_{-1.00}$ &  $13.3^{+2.4}_{-2.6}$ \\ 
$b$  &  $0.122^{+0.090}_{-0.082}$ &  $0.116^{+0.096}_{-0.083}$ &  $0.27^{+0.19}_{-0.18}$ &  $0.27 \pm 0.18$ \\ 
$e\sin\omega$  & 0 (fixed) &  $-0.02^{+0.21}_{-0.17}$ & 0 (fixed) &  $0.22^{+0.16}_{-0.13}$ \\ 
$e\cos\omega$  & 0 (fixed) &  $-0.0025^{+0.0042}_{-0.0040}$ & 0 (fixed) &  $-0.29^{+0.12}_{-0.11}$ \\ 
$K$ (km s$^{-1}$)  &  $0.178^{+0.038}_{-0.039}$ &  $0.174^{+0.040}_{-0.041}$ &  $0.0137^{+0.0023}_{-0.0021}$ &  $0.0188 \pm 0.0026$ \\ 
$\gamma_{\mathrm{FIES}}$ (km s$^{-1}$)  &  $29.070 \pm 0.025$ &  $29.067^{+0.026}_{-0.027}$ &  $-0.0137 \pm 0.0032$ &  $-0.0147^{+0.0032}_{-0.0031}$ \\ 
$\gamma_{\mathrm{HARPS-N}}$ (km s$^{-1}$)  &  &  &  $3.3323 \pm 0.0044$ &  $3.3356^{+0.0018}_{-0.0024}$ \\ 
$\gamma_{\mathrm{HARPS}}$ (km s$^{-1}$)  &  &  &  $3.3428 \pm 0.0017$ &  $3.3411^{+0.0021}_{-0.0023}$ \\
jitter$_{FIES}$ (km s$^{-1}$)  &  $0.027^{+0.030}_{-0.019}$ &  $0.028^{+0.029}_{-0.019}$ &  $0.0042^{+0.0047}_{-0.0029}$ &  $0.0032^{+0.0037}_{-0.0022}$ \\ 
jitter$_{HARPS-N}$ (km s$^{-1}$)  &  &  &  $0.0121^{+0.0054}_{-0.0033}$ &  $0.0046^{+0.0034}_{-0.0021}$ \\ 
jitter$_{HARPS}$ (km s$^{-1}$)  &  &  &  $0.0045^{+0.0020}_{-0.0013}$ &  $0.0057^{+0.0027}_{-0.0019}$ \\

		\hline
        \multicolumn{5}{l}{Derived Parameters} \\
       
$\delta$ (\%)  &  $0.9470^{+0.0062}_{-0.0048}$ &  $0.9470^{+0.0064}_{-0.0053}$ &  $0.2789^{+0.0111}_{-0.0054}$ &  $0.2801^{+0.0102}_{-0.0054}$ \\ 
$i$ ($^{\circ}$)  &  $88.67^{+0.89}_{-1.00}$ &  $88.76^{+0.86}_{-1.00}$ &  $89.15^{+0.58}_{-0.75}$ &  $88.4^{+1.1}_{-1.9}$ \\ 
$\rho_{\star}$ ($\rho_{\odot}$)  &  $0.2881^{+0.0054}_{-0.0100}$ &  $0.33^{+0.60}_{-0.23}$ &  $0.558^{+0.055}_{-0.100}$ &  $0.100^{+0.160}_{-0.073}$ \\ 
$T_{14}$ (days)  &  $0.17365^{+0.00045}_{-0.00040}$ &  $0.171^{+0.041}_{-0.026}$ &  $0.2124^{+0.0018}_{-0.0014}$ &  $0.213^{+0.057}_{-0.029}$ \\ 
$T_{23}$ (days)  &  $0.14196^{+0.00049}_{-0.00073}$ &  $0.140^{+0.033}_{-0.021}$ &  $0.1891^{+0.0015}_{-0.0027}$ &  $0.189^{+0.051}_{-0.025}$ \\ 
$T_{\mathrm{peri}}$ (BJD)  &  $2457820.738135 \pm 0.000090$ &  $2457820.757^{+1.281}_{-0.050}$ &  $2457906.84115 \pm 0.00045$ &  $2457907.58^{+0.59}_{-0.40}$ \\ 
$e$  & 0 (fixed) &  $0.134^{+0.124}_{-0.092}$ & 0 (fixed) &  $0.39 \pm 0.15$ \\ 
$\omega$ ($^{\circ}$)  & 90 (fixed) &  $262.0^{+7.9}_{-170.0}$ & 90 (fixed) &  $143.0 \pm 18.0$ \\ 
$a$ (AU)  &  $0.0404^{+0.0013}_{-0.0016}$ &  $0.0407^{+0.0071}_{-0.0078}$ &  $0.1365^{+0.0055}_{-0.0100}$ &  $0.102^{+0.019}_{-0.020}$ \\ 
$R_P$ ($R_J$)  &  $1.552^{+0.048}_{-0.057}$ &  $1.552^{+0.048}_{-0.057}$ &  $0.848^{+0.026}_{-0.022}$ &  $0.850^{+0.026}_{-0.022}$ \\ 
$M_P$ ($M_J$)  &  $1.42^{+0.31}_{-0.32}$ &  $1.39^{+0.32}_{-0.34}$ &  $0.163^{+0.028}_{-0.025}$ &  $0.223 \pm 0.031$ \\ 
$M_P\sin i$ ($M_J$)  &  $1.42^{+0.31}_{-0.32}$ &  $1.39^{+0.32}_{-0.34}$ &  $0.163^{+0.028}_{-0.025}$ &  $0.223 \pm 0.031$ \\ 
$M_P/M_{\star}$  &  $0.00107 \pm 0.00025$ &  $0.00104^{+0.00026}_{-0.00027}$ &  $0.000141^{+0.000024}_{-0.000022}$ &  $0.000194 \pm 0.000027$ \\ 
$\log g_P$ (cgs)  &  $3.164^{+0.091}_{-0.100}$ &  $3.154^{+0.095}_{-0.100}$ &  $2.750^{+0.071}_{-0.079}$ &  $2.884^{+0.059}_{-0.072}$ \\ 
$\rho_P$ (g cm$^{-3}$)  &  $0.50 \pm 0.12$ &  $0.49 \pm 0.13$ &  $0.355^{+0.067}_{-0.064}$ &  $0.483^{+0.076}_{-0.081}$ \\ 
$T_{\mathrm{eq}}$ (K)  &  $1957^{+78}_{-77}$ &  $1950^{+200}_{-190}$ &  $928^{+37}_{-19}$ &  $1080^{+110}_{-100}$ \\ 
$H$ (km)  &  $550 \pm 130$ &  $560 \pm 150$ &  $680 \pm 120$ &  $580^{+110}_{-100}$ \\ 
$2H/R_P$  &  $0.0102 \pm 0.0024$ &  $0.0104^{+0.0028}_{-0.0027}$ &  $0.0230^{+0.0040}_{-0.0042}$ &  $0.0195^{+0.0036}_{-0.0034}$ \\

        \hline
        \multicolumn{5}{p{500pt}}{Notes: in the case of circular orbits the epoch of periastron $T_{\mathrm{peri}}$ is identical to the transit epoch as we fix $\omega=90^{\circ}$. The atmospheric scale height $H$ assumes a hydrogen-dominated atmosphere (i.e., a mean molecular weight of 2), and the quantity $2H/R_P$ is the fractional surface area of the planetary disk subtended by an atmosphere of thickness $H$.  }
	\end{tabular}
\end{table*}

\subsection{Stellar rotation and inclination}
\label{sec:stellarrotation}

Stellar variability with a quasi-sinusoidal shape is easily visible in the {\it K2} light curve of \epichj\ (Fig.~\ref{fig:K2LCs}). The period of the variations, approximately 2 days, is too long to be due to $\delta$ Sct pulsations, and $\gamma$ Dor pulsators tend to be hotter than \epichj\ is \citep[see e.g.][]{Uytterhoeven11}. This may therefore be rotational variability. We performed a Fourier analysis of the light curve, finding that the most dominant frequency is at 0.462 cycles day$^{-1}$ ($P=2.16$ days); the first harmonic at 0.92 cycles day$^{-1}$ ($P=1.08$ days) is also present. This frequency structure is what would be expected for rotation with a period near 2 days; pulsations would result in a different frequency structure. We cannot, however, exclude the possibility that the 2-day peak could in fact be a harmonic, and that the actual rotation period could be an integer multiple of this. We will discuss this possibility in more detail below. 
There are also several strong peaks with periods of ten to several tens of days, which could be due to evolution of the spot pattern, but a detailed frequency analysis is beyond the scope of this paper. 

We measured the details of the 2-day peak using two different methodologies: a Lomb-Scargle periodogram analysis \citep{Lomb76,Scargle82,ZechmeisterKurster09}, and an autocorrelation function analysis \citep{McQuillan14}. In both cases we estimated the uncertainty on the period using the FWHM of the relevant peak. From the periodogram analysis we found a period of $2.169 \pm 0.048$ days, and from the autocorrelation function $2.16 \pm 0.49$ days. Both values are consistent but we conservatively adopt the less-certain autocorrelation value. 

As mentioned above, while the 2.16-day period may be the rotation period of \epichj, we cannot exclude the possibility that the rotation period may in fact be an integer multiple of this value. In Table~\ref{tab:rotpars} we list the rotation period, expected equatorial rotation velocity, implied value of $\sin i_{\star}$, and 1$\sigma$ allowed range of $i_{\star}$ assuming that the true rotation period is 1, 2, 3, or 4 times the 2.16-day period from the light curve.

Several points of interest are revealed by this analysis. First, if the rotation period is in fact 2.16 days, the predicted equatorial rotational velocity is $39.6 \pm 9.0$ \kms, more than twice the measured \vsinistar\ of $16.0 \pm 2.0$ \kms. This would require that the rotation axis of \epichj\ be significantly inclined with respect to the plane of the sky: $i_{\star}=23.8_{-6.4}^{+6.7\circ}$. An equatorial velocity of $40$ \kms, however, is much higher than typical for stars with the spectral type of \epichj; stars with \teff$\sim6350$ K typically have \vsinistar$<20$ \kms\ \citep[see e.g. Fig.~4 of][]{Winn17}. This suggests that it is {\it a priori} unlikely that the rotation period of \epichj\ is in fact 2.16 days, but this cannot be excluded on the basis of our current data.

If the rotation period of \epichj\ is in fact two or three times this (4.32 or 6.5 days), then the expected equatorial velocities are $19.8 \pm 4.5$ or $13.2 \pm 3.0$ \kms, respectively, both of which are consistent to within $1\sigma$ with the measured \vsinistar, and are also well within the range of \vsinistar\ values expected for a star of the spectral type of \epichj. On the other hand, in order to explain the quasi-sinusoidal shape of the light curve on a 2.16-day period with a longer rotation period, the spot distribution on the surface of \epichj\ would need to have a near-symmetrical two- or three-spot configuration. In these cases, a large inclination of the stellar rotation axis is not required. Indeed, we are only able to set loose 1$\sigma$ constraints of $i_{\star}>36.8^{\circ}$ and $>63.7^{\circ}$ for these two cases. For the 6.5-day rotation period case we note that this requires $\sin i_{\star}>1$ (we find $\sin i_{\star}=1.22 \pm 0.32$), but this is still consistent with 1 to within 1$\sigma$. A rotation period of 4 times 2.16 days or greater would require a value of $\sin i_{\star}$ increasingly larger than 1, excluding such values. These would also require increasingly complex and near-symmetrical multi-polar distributions of spots, which are unlikely. 

\begin{table}
	\centering
	\caption{Possible rotation periods of \epichj\ and consequences thereof}
	\label{tab:rotpars}
	\begin{tabular}{lcccc} 
		\hline
		 & $P_{\mathrm{rot}}$ & $v_{\mathrm{eq}}$ & $\sin i_{\star}$ & 1$\sigma$ range $i_{\star}$  \\
         & (days) & (\kms) & & ($^{\circ}$) \\
		\hline
	1 & $2.16 \pm 0.49$ & $39.6 \pm 9.0$ & $0.40 \pm 0.10$ & $17.4-30.6$ \\
    2 & $4.32 \pm 0.98$ & $19.8 \pm 4.5$ & $0.81 \pm 0.21$ & $>36.8$ \\
    3 & $6.5 \pm 1.5$ & $13.2 \pm 3.0$ & $1.22 \pm 0.32$ & $>63.7$ \\
    4 & $8.6 \pm 2.0$ & $9.9 \pm 2.3$ & $1.61 \pm 0.43$ & none \\
    
        \hline
        
	\end{tabular}
\end{table}

We therefore conclude that the rotation period of \epichj\ is one of 2.16, 4.32, or 6.5 days, but cannot confidently distinguish among these possibilities. We will discuss the implications of these possible rotation periods in more detail in \S\ref{sec:spinorbit}. 

Although sufficient spot coverage to induce measurable rotational modulation is unusual for late F stars, it is not unprecedented; for instance, \cite{Mazeh15} were able to measure rotation periods for numerous stars in the {\it Kepler} field with \teff\ values as high as 6500~K. 

We did not detect any significant rotational modulation in the light curve of \epicws, and so we cannot perform a similar analysis for that star. Based upon our values of \vsinistar\ and \rstar\ for this object, we calculate a $1\sigma$ upper limit on the expected rotation period of $37.2$ days. The more pole-on \epicws\ is viewed, the shorter the rotation period must be to explain the measured \vsinistar.

\subsection{Secondary eclipse of \epichj}
\label{sec:2ndaryeclipse}

\begin{figure}
	\includegraphics[width=\columnwidth]{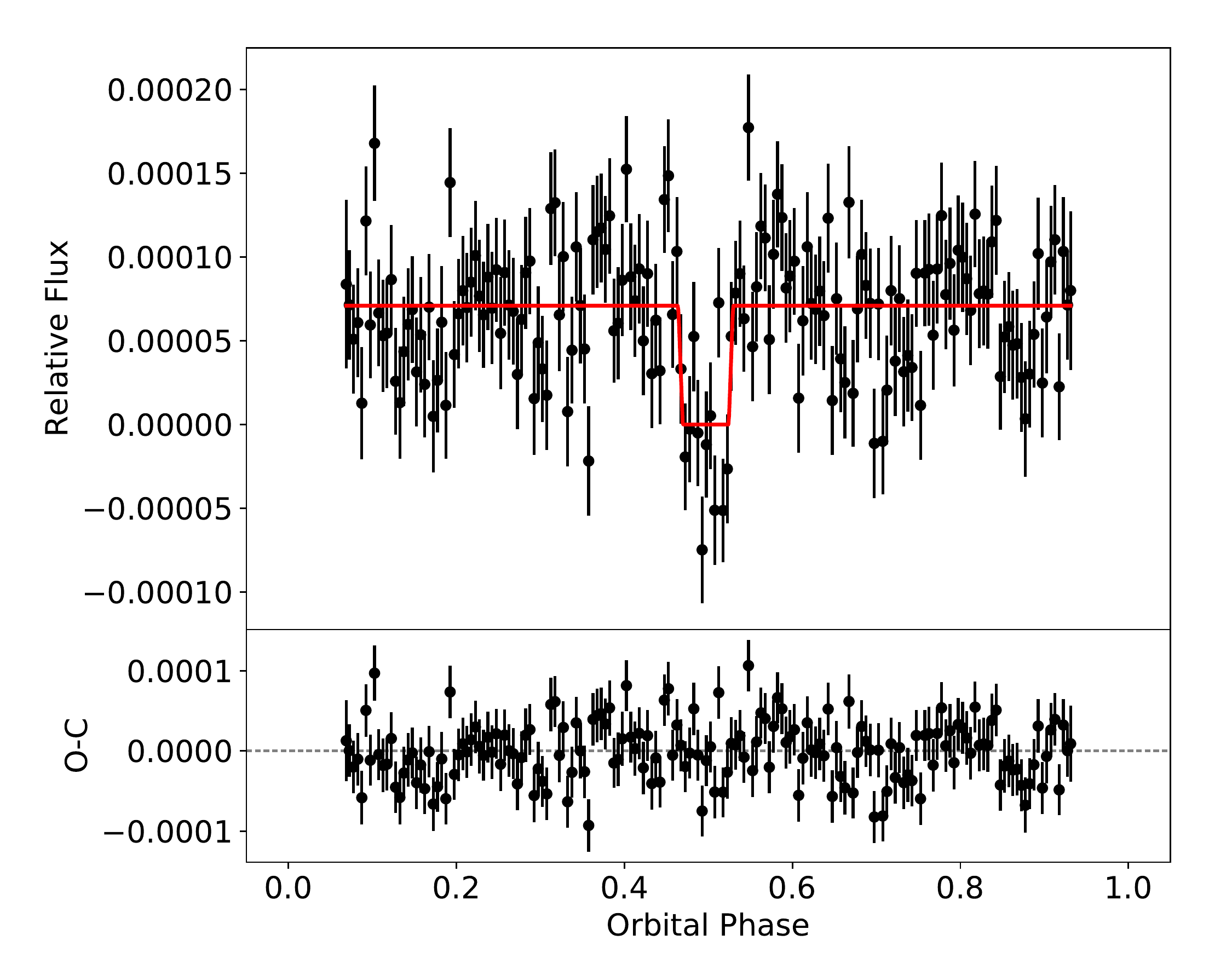}\\
    \caption{{\it K2} light curve of \epichj\ (top), after the stellar variability and in-transit data have been removed and the data have been folded and binned on the orbital period. We show the data in black, and the best-fit secondary eclipse model as a dashed line (red online). The residuals to the fit are shown in the bottom panel.}
    \label{fig:phasecurve}
\end{figure}

We searched the {\it K2} light curve for a secondary eclipse for \epichj, which, if of large enough magnitude, could have indicated that the system was a false positive. We subtracted the stellar variability of \epichj\ by fitting a second-order polynomial to each 0.25$\times P_{\mathrm{orb}}$ portion of the out-of-transit light curve. This aggressive detrending procedure was necessitated by the stellar variability with a period close to the planetary orbital period (2.16 days versus 2.63 days; see \S\ref{sec:stellarrotation}). Although this detrending should have also removed any phase curve which might be present in the {\it K2} light curve, disentangling the phase curve and stellar variability will be very difficult (see \S\ref{sec:eclipsediscuss} for further discussion), and so we chose to detrend out both the stellar variability and the phase curve and concentrate on detecting the secondary eclipse. We phase-folded and binned the detrended {\it K2} light curve on the planetary orbital period, which is shown in Fig.~\ref{fig:phasecurve}. We do indeed detect a secondary eclipse. 

We fit the eclipse with an occultation model and an MCMC procedure. We obtain a best-fit secondary eclipse depth of $\delta_{\mathrm{sec}}=71 \pm 15$ ppm, as well as $e\cos\omega=-0.0049_{-0.0036}^{+0.0048}$. The secondary eclipse depth is small, and is consistent with \epichj~b being a planet (which is also confirmed by our RV measurements). See \S\ref{sec:eclipsediscuss} for further discussion.

The value of $e\cos\omega$ that we obtained is consistent with zero to within $1.4\sigma$. We thus conclude that there is no compelling evidence from the secondary eclipse for any orbital eccentricity, nor is there from the radial velocity data (\S\ref{sec:fitting}).

\section{Discussion}

\subsection{Properties of the planets and systems}
\label{sec:planetprops}

\begin{figure}
	\includegraphics[width=\columnwidth]{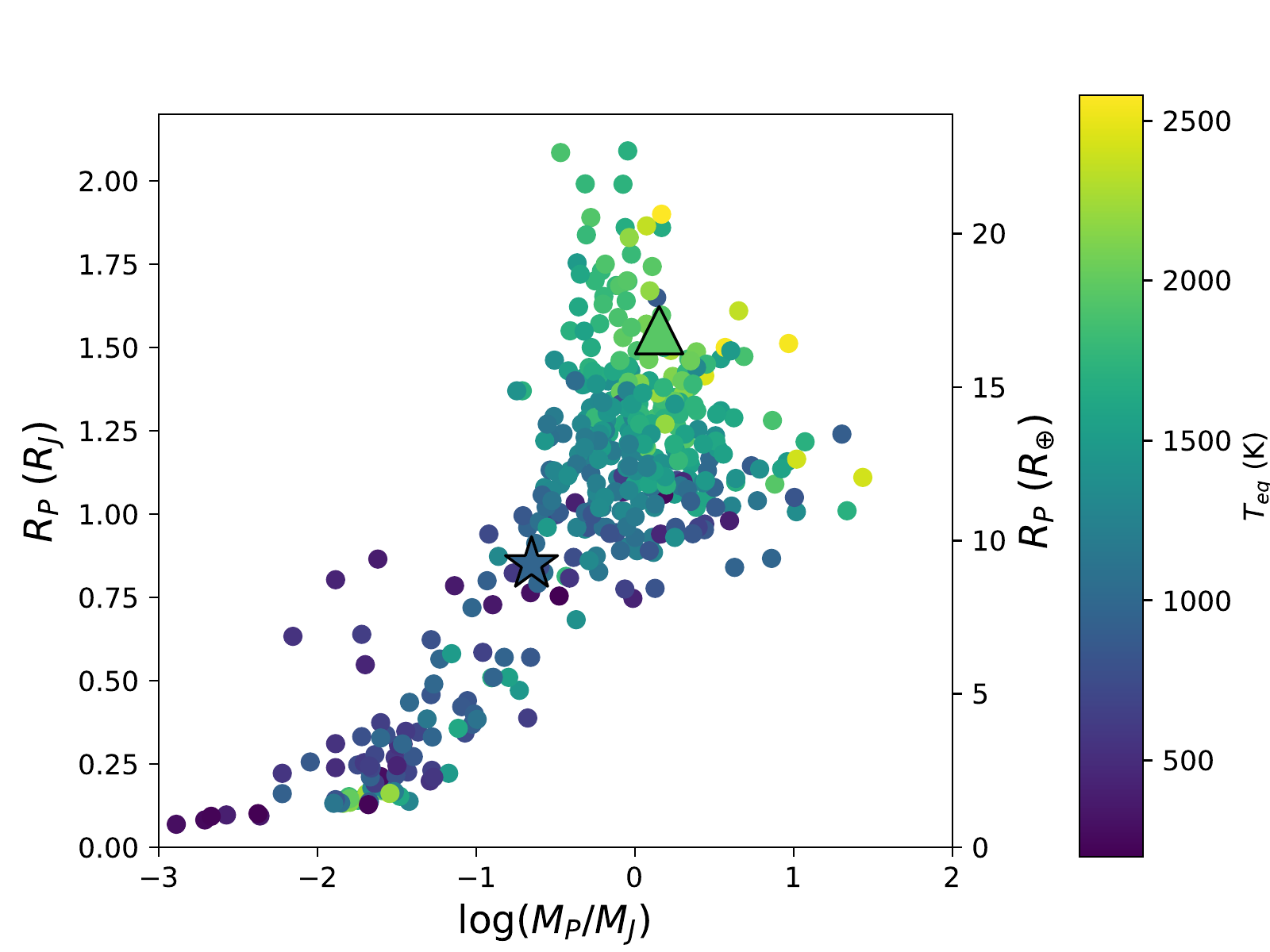}\\
    \caption{\epichj~b (triangle) and \epicws~b (star) in context with the population of transiting planets with mass and radius measurements, as obtained from the NASA Exoplanet Archive (\url{https://exoplanetarchive.ipac.caltech.edu/}) on 2018 March 7. 
    The plot points are coloured according to the zero-albedo equilibrium temperature $T_{\mathrm{eq}}$.}
    \label{fig:MRdiag}
\end{figure}

In many ways both \epichj~b and \epicws~b are typical representatives of the populations to which they belong. In Fig.~\ref{fig:MRdiag} we show these two planets in context in the mass-radius diagram.

\epichj~b is a fairly typical hot Jupiter of 1.4 \mj\ and 1.6 \rj\ on a circular 2.6-day orbit. It is slightly inflated, which is expected for its mass and zero-albedo equilibrium temperature of $\sim2000$ K; planets in this region of the mass-radius diagram tend to have similar equilibrium temperatures (Fig.~\ref{fig:MRdiag}). \epichj~b is most noteworthy in that there is a possibility that its orbit may be misaligned with respect to its host star's rotation (\S\ref{sec:spinorbit}), and that it shows a secondary eclipse in the \kepler\ bandpass (\S\ref{sec:eclipsediscuss}).

\epicws~b is also exemplary of the population of warm Saturns. Although its radius is somewhat too large to be part of the population of sub-Saturns identified by \cite{Petigura17}--its radius is 9.2 \re, while \cite{Petigura17} defined sub-Saturns as having radii between 4 and 8 \re--it nonetheless follows many of the same trends seen in this population. For instance, like warm sub-Saturns around metal-rich stars \citep{Petigura17}, \epicws~b has an eccentric orbit, no additional known planets in the system, and a relatively high mass ($65\,M_{\oplus}$, which is at the upper end of the range typical of sub-Saturns).

\epicws\ is also interesting in that it is one of the older planet-host stars with a well-determined age. The NASA Exoplanet Archive\footnote{\url{https://exoplanetarchive.ipac.caltech.edu/}} lists only eighteen planet-host stars (as of 2017 May 24) with ages of $>8.5$ Gyr and an age uncertainty of less than 2 Gyr. \epicws\ is among the most metal-rich of these systems, although its metallicity is not unprecedented for a star of this age \citep[cf. Fig.~10 of][]{SilvaAguirre18}.

\epicws~b is additionally one of three recently discovered $P\sim12$ day warm Saturns on eccentric orbits around bright stars from {\it K2}, along with K2-232~b \citep{Brahm18,Yu18} and K2-234~b \citep{VanEylen18,Yu18}. Although \epicws~b is somewhat fainter than either of these stars (which have $V=9.3$ and $V=9.9$ respectively), it is nonetheless bright enough to enable many follow-up observations, and will allow comparisons to these other planets. It is intermediate in mass and radius to these two, and the system also has a very similar age to K2-234 \citep{VanEylen18}, which will also be helpful for comparative planetology. 

\subsection{Candidate Stellar Companion to \epichj}
\label{sec:companion}

As discussed in \S\ref{sec:AO}, we detected a candidate stellar companion to \epichj\ using our Subaru IRCS high-resolution imaging. As can be seen in Fig.~\ref{fig:Subaru}, the candidate companion is close to the speckle pattern, but we found that it has a contrast of more than $5\sigma$ and is therefore likely to be real.

As we only have a single epoch of single-band imaging, we cannot confirm the association of the candidate companion via the usual methods of common proper motion or colours indicating that it is consistent with being located at the same distance. We therefore must rely upon probabilistic arguments. In order to assess the probability that the candidate companion is indeed physically associated with \epichj, we followed essentially the same methodology as used by \cite{Johnson18} to estimate the probability that the candidate companion could be a background source. We obtained Galactic models computed by the \textsc{trilegal}\footnote{\url{http://stev.oapd.inaf.it/cgi-bin/trilegal\_1.6}} \citep{Girardi05} and \textsc{Besan\c{c}on}\footnote{\url{http://model2016.obs-besancon.fr/}} \citep{Robin03} codes for the coordinates of \epichj. We used the default parameters for both codes, except for using the \cite{Schlafly11} reddening maps\footnote{\url{https://irsa.ipac.caltech.edu/applications/DUST/}} for \textsc{trilegal} and the \cite{Marshall06} dust map for \textsc{Besan\c{c}on}. We then counted the number of sources in each model, and used this to estimate the probability that there would be a background source at least as bright as the candidate companion at least as close to \epichj\ as the candidate companion. We found that the probability of such a chance superposition is 0.13\% for the \textsc{trilegal} model, and 0.016\% for the \textsc{Besan\c{c}on} model. Although these estimates are in disagreement with each other by about an order of magnitude, we nonetheless conclude that the candidate companion is very likely to be physically associated with the \epichj\ system. We caution, however, that we cannot confirm its association with our current data. We proceed for the remainder of the section with further analyses under the assumption that the candidate companion is indeed physically bound to \epichj.

We used the \textsc{isochrones} package \citep{Morton16} to estimate what stellar properties of the candidate companion would result in the observed $\Delta H$ value given the properties of the primary. Using the 2MASS $H$-band magnitude and Gaia DR2 parallax, \epichj\ has an absolute magnitude of $M_H=2.039 \pm 0.069$, and the candidate companion therefore has $M_H=8.78 \pm 0.19$. This corresponds to a mass of $\sim0.13-0.16$\msun\ for the candidate companion, taking into account the uncertainties on the age and metallicity of the primary. This would correspond to \teff$\sim3200-3300$~K, or, per \cite{KrausHillenbrand07}, a spectral type of M5-6V.

The candidate companion is too faint to significantly affect our other analyses of this system. Assuming the stellar parameters found earlier, the companion would have a magnitude ratio in the \kepler\ bandpass of $\Delta Kp\sim9.5$ (again using \textsc{isochrones}). This would result in a dilution of the transit depth of \epichj\ of less than 2 parts in $10^{-5}$, which is much smaller than the uncertainty on the transit depth due to the photometric noise in the light curve, and is thus negligible. Similarly, the companion is too faint to be an eclipsing binary causing the transits of \epichj; in order to cause a transit of the observed depth, even a 100 per cent deep eclipse would need to occur on a companion no fainter than $\Delta Kp\sim5.1$. This possibility is also excluded by our detection of the reflex motion of \epichj\ due to the planet, and the achromaticity of the transit in our MuSCAT2 data.

The discovery of a candidate stellar companion to \epichj\ is scientifically interesting for a number of reasons. \cite{Ngo16} found that host stars of hot Jupiters are significantly more likely than field stars to host stellar companions between 50 and 2000 AU, which tend to have mass ratios smaller than would be expected given the population of field binaries. The candidate companion to \epichj\ fits this trend, with its projected separation of 400 AU and mass ratio of $\sim0.1$, and is furthermore close and massive enough to likely be able to drive planetary migration via Kozai-Lidov oscillations \citep[e.g.,][]{FabryckyTremaine07,Naoz12} We therefore encourage additional observations of \epichj\ to confirm whether the candidate companion is bound to the system.

\subsection{Orbital alignment of \epichj~b}
\label{sec:spinorbit}

Based upon the rotational variability in the light curve of \epichj\ (\S\ref{sec:stellarrotation}), we found that the rotation axis of \epichj\ may be significantly inclined with respect to the plane of the sky, depending upon whether the true rotation period is 2.16 days or an integer multiple thereof. This has consequences for the orbit of the planet. If the stellar rotation axis is inclined with respect to the plane of the sky, then the planetary orbit must have a substantial obliquity with respect to the stellar spin in order to transit. This would not be unexpected, as \epichj\ is above the Kraft break \citep{Kraft67}, the point on the main sequence above which stars rotate rapidly due to their lack of a deep surface convective zone, a strong magnetic dynamo, and consequent rotational braking. Hot Jupiters around stars above the Kraft break often have substantially misaligned orbits \citep{Winn10,Albrecht12}. If instead the rotation period is twice or three times 2.16 days, a large stellar inclination, and therefore a misaligned orbit, are not required. This does not, however, exclude the possibility of a misaligned orbit; even if the stellar rotation axis is perpendicular to the line of sight, the sky-projected spin-orbit misalignment $\lambda$ can be large. We cannot constrain $\lambda$ with our current data, but this could easily be measured using the Rossiter-McLaughlin effect (\S\ref{sec:futureobs}). 

We also note that if the rotation period is in fact 2.16 days, this is close to the 2.63-day planetary orbital period. Systems with $P_{\mathrm{orb}}\sim P_{\mathrm{rot}}$ have different tidal dynamics from those with rotation periods much different from the orbital period, including likely enhanced tidal dissipation \citep[e.g.,][]{CollierCameronJardine18}, and therefore \epichj\ may be of interest from this standpoint. As also discussed by \cite{CollierCameronJardine18}, the dissipation is likely different for spin-orbit misaligned systems near the stellar rotation period due to the different motion of the tidal bulge, providing additional motivation to measure the Rossiter-McLaughlin effect for \epichj\ (\S\ref{sec:futureobs}). In addition, \epichj\ may join a number of F stars with hot Jupiters where the stellar rotation is quasi-synchronized to the planetary orbit, such as $\tau$ Boo \citep{Donati08} and CoRoT-4 \citep{Lanza09}. Such systems were predicted to be an effective end-point of hot Jupiter tidal evolution for F stars by \cite{DamianiLanza15}, as reduced tidal damping when $P_{\mathrm{rot}}\sim P_{\mathrm{orb}}$ will delay the planet reaching the point where it can be tidally disrupted. \epichj's relatively advanced age for an F star, near the main-sequence turn-off, is consistent with this picture; the system has had ample time to evolve into such a quais-stable state. On the other hand, \cite{CollierCameronJardine18} found evidence of {\it stronger} tidal damping for (aligned) systems with $P_{\mathrm{rot}}\sim P_{\mathrm{orb}}$. \epichj\ could thus offer an opportunity to test the impact of any spin-orbit misalignment upon this evolution.

\subsection{Secondary eclipse of \epichj~b}
\label{sec:eclipsediscuss}

\epichj~b is one of a relatively small number of planets with a significantly detected secondary eclipse in the \kepler\ bandpass, and the first such hot Jupiter discovered using {\it K2} \citep[although K2-113~b has a $1.9\sigma$ detection of the secondary eclipse, and the ultra short period planet K2-141~b also has a detected secondary eclipse:][]{Espinoza17,Malavolta18}.

From \cite{Esteves13}, the secondary eclipse depth in the \kepler\ bandpass is
\begin{equation}
\label{eqn:eclipsedepth}
\delta_{\mathrm{sec}}=\bigg(\frac{R_P}{\rstar}\bigg)^2\frac{\int B_{\lambda}(T_B)T_K d\lambda}{\int F_{\lambda} T_K d\lambda}+A_g\bigg(\frac{R_P}{\rstar}\frac{\rstar}{a}\bigg)^2
\end{equation}
where $B_{\lambda}$ is the Planck function, $T_B$ is the blackbody temperature of the planetary dayside, $T_K$ is the \kepler\ bandpass transmission function, $F_{\lambda}$ is the stellar flux, and $A_g$ is the planetary geometric albedo. This is Eqn.~18 of \cite{Esteves13}, slightly rewritten to accommodate the notation and measured quantities in this article. The first term on the right hand side of Eqn.~\ref{eqn:eclipsedepth} describes the thermal emission in the \kepler\ bandpass, while the second term is the contribution of reflected starlight.

We numerically solve this equation for the geometric albedo $A_g$, assuming, after \cite{Esteves13}, that $T_B=T_{\mathrm{eq}}(A_B)$, where $A_B=3/2A_g$ is the Bond albedo, and $T_{\mathrm{eq}}=\teff\sqrt{f\rstar/a}(1-A_B)^{1/4}$, where $f$ describes the efficiency of heat redistribution. We used two different limiting cases: homogeneous re-distribution of heat ($f=1/4$) and instant re-radiation from the dayside ($f=2/3$). We also make the simplifying assumptions that the planetary emission and stellar flux are both blackbodies, and that the \kepler\ bandpass is a tophat function between 4000 \AA\ and 9000 \AA\ \citep[following][]{Huber17}. In both heat redistribution cases the eclipse depth requires a geometric albedo of $A_g\sim0.2$, and the contribution of thermal emission in the \kepler\ bandpass is small. Although this estimate is not likely to be completely accurate due to the approximations that we have made, it nonetheless suggests that \epichj~b is likely among the class of more reflective hot Jupiters \citep[e.g.,][]{HengDemory13,SheetsDeming17}.

The detection of a secondary eclipse that is dominated by reflected starlight also implies that there should in principle be a detectable phase curve in the optical. 
Our aggressive detrending procedure (\S\ref{sec:2ndaryeclipse}) should have removed the phase curve from our phase-folded light curve, and indeed no phase variation is visible in Fig.~\ref{fig:phasecurve}. The detection of the phase curve, however, would be difficult, as it would require careful detrending of the stellar variability. This would be particularly complicated as the 2.16-day period that dominates the stellar variability (\S\ref{sec:stellarrotation}) is similar to the 2.63-day orbital period (which is why we used an aggressive detrending procedure in the first place). Such a detailed search for the phase curve is beyond the scope of the present work.

\subsection{Potential for further observations}
\label{sec:futureobs}

\begin{figure}
	\includegraphics[width=1.0\columnwidth]{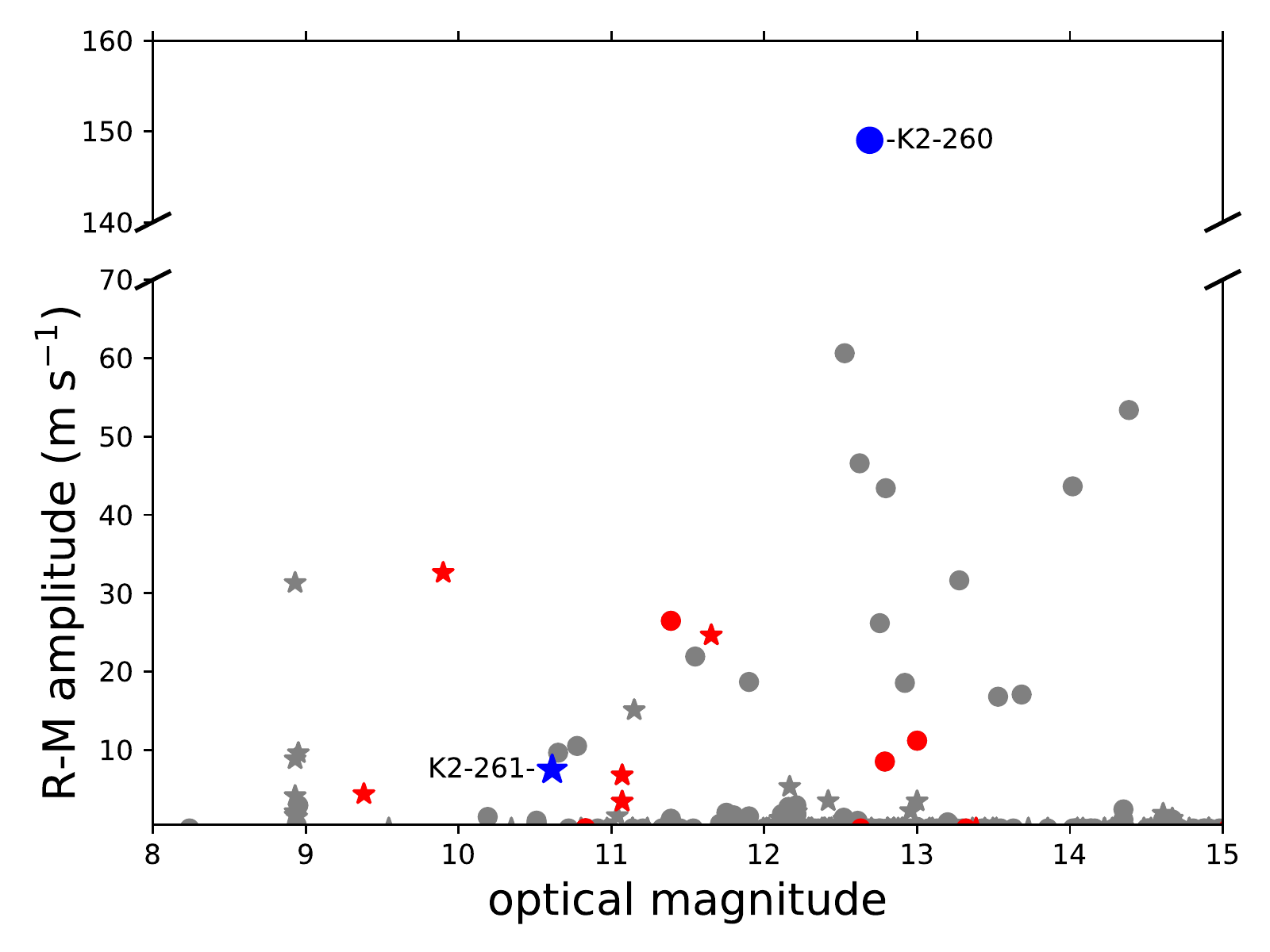}\\
    \caption{Predicted Rossiter-McLaughlin amplitude as a function of optical magnitude (either $V$ or $Kp$ depending upon the target) for transiting planets discovered by or observed by {\it K2}. Approximately Saturn-sized planets ($0.5$\rj$<R_P<0.9$\rj) are highlighted in black (red online), while \epichj\ and \epicws\ are shown in black (blue online) and are noted on the plot. Planets with $P<10$ days are depicted as circles, while those with $P\geq10$ days are depicted as stars. \epicws\ is among the best Saturns and sub-Saturns, and one of the better long-period planets, for an obliquity measurement due to its combination of relatively large Rossiter-McLaughlin amplitude and bright host star. The axes are broken in order to better show \epichj\ along with the rest of the population; no {\it K2} planets in the depicted magnitude range have expected R-M amplitudes between 70 and 140 \ms, and so the missing region of the plot is empty.}
    \label{fig:RMamps}
\end{figure}

Both \epichj~b and \epicws~b are promising targets for future observations to further characterize these systems. As discussed in \S\ref{sec:spinorbit}, 
it is possible that \epichj~b has an orbit misaligned with respect to the stellar spin axis. 
Measuring the sky-projected component of this misalignment ($\lambda$) should be straightforward with either Rossiter-McLaughlin \citep[e.g.,][]{Triaud10} or Doppler tomographic \citep[e.g.,][]{DT3} observations. Given the properties of this system, we predict a Rossiter-McLaughlin semi-amplitude of $\sim150$ \ms\ for \epichj~b using the formulae of \cite{GaudiWinn07}. \epicws~b is also a promising target for such observations, with a predicted Rossiter-McLaughlin amplitude of $\sim7.5$ \ms. Furthermore, the transit duration is relatively short for a long-period planet ($\sim5$ hours), allowing a full transit to be observed in a single night, which is critical for RV Rossiter-McLaughlin observations. \epichj~b and \epicws~b are among the better {\it K2} targets for such observations; indeed, \epichj~b has the highest expected Rossiter-McLaughlin amplitude of any confirmed or validated {\it K2} planet orbiting a star brighter than $V=15$. We show these objects' expected Rossiter-McLaughlin amplitudes in context with the remainder of the {\it K2} planet population in Fig.~\ref{fig:RMamps}. 
These will require the use of large telescopes. As can be seen in Table~\ref{tab:RVs}, for \epichj\ with FIES we are only able to obtain an RV precision of $\sim60-150$ \ms with 1-hour exposures, which is insufficient to properly resolve the Rossiter-McLaughlin effect; ideally one would need at least 8-10 exposures over the 4-hour transit duration. For \epicws\ we are able to obtain a per-point uncertainty of $\sim1-3$ \ms with $\sim1200-1800$ s exposures with HARPS or HARPS-N, which should be adequate to detect the Rossiter-McLaughlin effect; again, with FIES we are only able to obtain a per-point precision of $\sim10$ \ms\ with 3600-second exposures, insufficient to detect the Rossiter-McLaughlin effect for this target.

\epicws\ is among the brightest stars to host a transiting giant planet with $P>10$ days, which will facilitate other follow-up observations, in particular atmospheric transmission spectroscopy. Thanks to its relatively high temperature ($T_{\mathrm{eq}}=1080^{+110}_{-100}$ K, assuming zero albedo and neglecting the orbital eccentricity) and low surface gravity ($\log g_P=2.884^{+0.059}_{-0.072}$), we estimate a planetary atmospheric scale height of $580^{+110}_{-100}$ km, assuming a hydrogen-dominated atmosphere. For \epichj~b we estimate a scale height of $550 \pm 130$ km, similar to \epicws~b as the higher planetary gravity largely offsets the higher equilibrium temperature. Although this is also reasonably favourable for transmission spectroscopy, at $V=12.7$ these observations would be challenging simply because of the relative faintness of the host star.

Long-term RV monitoring of both systems could find or constrain the presence of additional planets in these systems. As discussed in \S\ref{sec:fitting} we found some evidence for a possible RV trend for \epicws, but rejected this possibility as it shifted the parameters of the transiting planet to imply an unphysical solution. Further observations could determine if there in fact might be another planet in the system, either by uncovering a long-term RV trend or finding an RV periodicity due to a shorter-period planet. 
Our RV monitoring only spans $\sim100$ days for \epichj\ and $\sim130$ days for \epicws, insufficient to detect any small, long-term trends.

\section{Conclusions}

We have presented the discovery, confirmation, and initial characterization of two transiting planets from {\it K2}, \epichj~b and \epicws~b. 

\epichj~b is a somewhat inflated hot Jupiter on a 2.6-day orbit around a mildly rapidly rotating (\vsinistar$=16.0 \pm 2.0$ \kms) F6V star. 
We detect the secondary eclipse of the planet in the \kepler\ bandpass, which we find to be largely due to reflected starlight and use to estimate a planetary geometric albedo of $A_g\sim0.2$.  
The host star also exhibits rotational modulation in its lightcurve. If the dominant periodicity of 2.16 days is the stellar rotation period then the planetary orbit must be misaligned with respect to the stellar rotation axis, as the predicted equatorial rotational velocity is much larger than the measured \vsinistar; if, however, the 2.16 day periodicity is a harmonic of the true rotation period, then no large spin-orbit misalignment is required.

\epicws~b is a warm Saturn on an eccentric ($e=0.39$), 11.6-day orbit around a bright ($V=10.6$), metal-rich (\feh$=+0.36$) 8.8 Gyr old G7 star at the main sequence turn-off. The star is most likely a member of the Galactic thin disk, and is among the brightest stars to host a transiting giant planet with a period of greater than 10 days.  

\epichj~b and \epicws~b add to the sample of transiting giant planets discovered by {\it K2}, and are both promising targets for follow-up observations to further characterize these systems. These could include observations of the Rossiter-McLaughlin effect (indeed, \epichj~b should have the largest Rossiter-McLaughlin amplitude of any planet discovered with {\it K2}), atmospheric transmission spectroscopy, and long-term RV monitoring.

During the revision process of this paper we became aware of an independent discovery of \epicws~b by \cite{Brahm18-8078}. No information about this object, including on analysis and results, were shared between the two teams prior to the submission of either paper.

\section*{Acknowledgements}

We thank the anonymous referee for helpful comments which improved the quality of the paper.

F.D. and J.N.W. thank the Heising-Simons Foundation for financial support. %from Josh
Funding for the Stellar Astrophysics Centre is provided by The Danish National Research Foundation (Grant agreement no.: DNRF106). %from Anders
This research was partly supported by JSPS KAKENHI Grant Number JP18H01265 and JST PRESTO Grant Number JPMJPR1775, Japan. %from Norio
A.P.H., Sz.Cs., S.G., J.K., M.P., and H.R. acknowledge support by DFG grants HA 3279/12-1, PA525/18-1, PA525/19-1, PA525/20-1, and RA 714/14-1 within the DFG Schwerpunkt SPP 1992, ``Exploring the Diversity of Extrasolar Planets.'' %from Judith and Artie
W.D.C., M.E., and P.J.M. acknowledge support from NASA grants NNX16AJ11G and 80NSSC18K0447 to The University of Texas at Austin. %from Bill
M.F. and C.M.P. gratefully acknowledge the support of the Swedish National Space Board. %from Carina
This project has received funding from the European Union's Horizon 2020 research and innovation programme under grant agreement No 730890. This material reflects only the authors views and the Commission is not liable for any use that may be made of the information contained therein. %from Davide

We are very grateful to the McDonald, NOT, TNG, and ESO staff members for their support during the observations. This paper includes data taken at The McDonald Observatory of The University of Texas at Austin. Based on observations obtained \emph{a}) with the Nordic Optical Telescope (NOT), operated on the island of La Palma jointly by Denmark, Finland, Iceland, Norway, and Sweden, in the Spanish Observatorio del Roque de los Muchachos (ORM) of the Instituto de Astrofisica de Canarias (IAC), under programs 56-010, 56-112, and 56-209; \emph{b}) with the Italian Telescopio Nazionale Galileo (TNG) also operated at the ORM (IAC) on the island of La Palma by the INAF - Fundacion Galileo Galilei, under Spanish CAT program CAT17B\_99, OPTICON program OPT17B\_59, and program A36TAC\_12, \emph{c}) with the 3.6m ESO telescope at La Silla Observatory under program 0100.C-0808 and 0101.C-0829. 
Based in part on data collected at Subaru Telescope, which is operated by the National Astronomical Observatory of Japan. The authors wish to recognize and acknowledge the very significant cultural role and reverence that the summit of Mauna Kea has always had within the indigenous Hawaiian community.  We are most fortunate to have the opportunity to conduct observations from this mountain.
This article is partly based on observations made with the MuSCAT2 instrument, developed by ABC, at Telescopio Carlos S\'anchez operated on the island of Tenerife by the IAC in the Spanish Observatorio del Teide.
This work has made use of data from the European Space Agency (ESA) mission
{\it Gaia} (\url{https://www.cosmos.esa.int/gaia}), processed by the {\it Gaia}
Data Processing and Analysis Consortium (DPAC,
\url{https://www.cosmos.esa.int/web/gaia/dpac/consortium}). Funding for the DPAC
has been provided by national institutions, in particular the institutions
participating in the {\it Gaia} Multilateral Agreement.
This publication makes use of data products from the Wide-field Infrared Survey Explorer, which is a joint project of the University of California, Los Angeles, and the Jet Propulsion Laboratory/California Institute of Technology, funded by the National Aeronautics and Space Administration. This research made use of NASA's Astrophysics Data System and the NASA Exoplanet Archive, which is
operated by the California Institute of Technology, under contract with the National Aeronautics and Space Administration under the Exoplanet Exploration Program. This research made use of Astropy, a community-developed core Python package for Astronomy \citep{astropy}.

%%%%%%%%%%%%%%%%%%%%%%%%%%%%%%%%%%%%%%%%%%%%%%%%%%



\bibliographystyle{mnras}
\bibliography{bibmaster} % if your bibtex file is called example.bib

%\begin{thebibliography}{99}
%\end{thebibliography}

%%%%%%%%%%%%%%%%%%%%%%%%%%%%%%%%%%%%%%%%%%%%%%%%%%

%%%%%%%%%%%%%%%%% APPENDICES %%%%%%%%%%%%%%%%%%%%%

%\appendix



%%%%%%%%%%%%%%%%%%%%%%%%%%%%%%%%%%%%%%%%%%%%%%%%%%

% Don't change these lines
\bsp	% typesetting comment
\label{lastpage}
\end{document}